\documentclass[fleqn,usenatbib]{mnras}

\usepackage{newtxtext,newtxmath}
\usepackage[T1]{fontenc}

\DeclareRobustCommand{\VAN}[3]{#2}
\let\VANthebibliography\thebibliography
\def\thebibliography{\DeclareRobustCommand{\VAN}[3]{##3}\VANthebibliography}

\usepackage{graphicx}	% Including figure files
\usepackage{savesym}
\savesymbol{Bbbk}
\usepackage{amsmath}	% Advanced maths commands
\usepackage{amssymb}	% Extra maths symbols
\restoresymbol{Bbbk1}{Bbbk}
\usepackage{natbib}
\usepackage[flushleft]{threeparttable}
\usepackage{caption}
\usepackage{pdflscape}
\usepackage{color}
\usepackage{placeins}
\usepackage{capt-of}
\usepackage{soul}

\newcommand{\mum}{{$\mu$m\ }}
\newcommand{\um}{{$\mu$m}}
\newcommand{\dsec}{$\rlap{.}^s$}

\title[SCUBA-2 observations of Corona Australis]{The JCMT Gould Belt Survey: First results from the Corona Australis molecular cloud and evidence of variable dust emissivity indices in the Coronet region}
\author[K. Pattle et al.]{K. Pattle$^{1}$, D. Bresnahan$^{2}$, D. Ward-Thompson$^{2}$, H. Kirk$^{3}$, J. M. Kirk$^{2}$, D. S. Berry$^{4}$, \newauthor H. Broekhoven-Fiene$^{5}$, J. Hatchell$^{6}$, T. Jenness$^{7}$, D. Johnstone$^{4, 5}$, J. C. Mottram$^{8, 9}$, A. Duarte-Cabral$^{10}$, \newauthor J. Di Francesco$^{3, 5}$, M. R. Hogerheijde$^{8, 11}$, P. Bastien$^{12}$, H. Butner$^{13}$, M. Chen$^{14}$, A. Chrysostomou$^{15}$, \newauthor S. Coud\'{e}$^{16,17}$,  M. J. Currie$^{4,33}$, C. J. Davis$^{18}$, E. Drabek-Maunder$^{19}$, M. Fich$^{20}$, J. Fiege$^{21}$, P. Friberg$^{4}$, \newauthor R. Friesen$^{22}$, G.A. Fuller$^{23}$, S. Graves$^{4}$, J. Greaves$^{10}$, W. Holland$^{24, 25}$, G. Joncas$^{26}$, L.B.G. Knee$^{3}$,  \newauthor S. Mairs$^{3,4}$, K. Marsh$^{10}$, B. C. Matthews$^{3, 5}$, G. Moriarty-Schieven$^{3}$, C. Mowat$^{6}$, J. Rawlings$^{1}$, B. Retter$^{6}$, \newauthor  J. Richer$^{27, 28}$, D. Robertson$^{29}$, E. Rosolowsky$^{30}$, S. Sadavoy$^{14}$, H. Thomas$^{17}$, N. Tothill$^{31}$, S. Viti$^{8}$, \newauthor G. J. White$^{32, 33}$, J. Wouterloot$^{4}$, J. Yates$^{1}$, M. Zhu$^{34}$\\
\\
\textit{Affiliations are given after the references}}

\begin{document}

\date{}

\pagerange{\pageref{firstpage}--\pageref{lastpage}} \pubyear{2024}

\maketitle

\label{firstpage}

\begin{abstract}

We present 450\um\ and 850\um\ James Clerk Maxwell Telescope (JCMT) observations of the Corona Australis (CrA) molecular cloud taken as part of the JCMT Gould Belt Legacy Survey (GBLS).  We present a catalogue of 39 starless and protostellar sources, for which we determine source temperatures and masses using SCUBA-2 450\um/850\um\ flux density ratios for sources with reliable 450\um\ detections, and compare these to values determined using temperatures measured by the \emph{Herschel} Gould Belt Survey (HGBS).  In keeping with previous studies, we find that SCUBA-2 preferentially detects high-volume-density starless cores, which are most likely to be prestellar (gravitationally bound).  We do not observe any anti-correlation between temperature and volume density in the starless cores in our sample.  Finally, we combine our SCUBA-2 and \textit{Herschel} data to perform SED fitting from 160--850$\mu$m across the central Coronet region, thereby measuring dust temperature $T$, dust emissivity index $\beta$ and column density $N({\rm H}_2)$ across the Coronet.  
{We find that $\beta$ varies across the Coronet, particularly measuring $\beta = 1.55 \pm 0.35$ in the colder starless SMM-6 clump to the north of the B star R CrA.  This relatively low value of $\beta$ is suggestive of the presence of large dust grains in SMM-6, even when considering the effects of $T-\beta$ fitting degeneracy and $^{12}$CO contamination of SCUBA-2 850$\mu$m data on the measured $\beta$ values.}

\end{abstract}


\section{Introduction}

The Corona Australis molecular cloud (CrA) is a nearby molecular cloud which is forming low- to intermediate-mass stars \citep{wilking1985, wilking1986,nutter2005}, located $\sim$17$^\circ$ to the south of the Galactic Plane. The most well-studied part of the cloud is the Coronet \citep{taylorandstorey1984}, a young cluster of objects \citep[e.g.][]{esplin2022} which includes the Herbig Ae/Be variable stars R CrA and T CrA (e.g., \citealt{siciliaaguilar2013}).  

\begin{figure*}
\centering
\includegraphics[width=\textwidth]{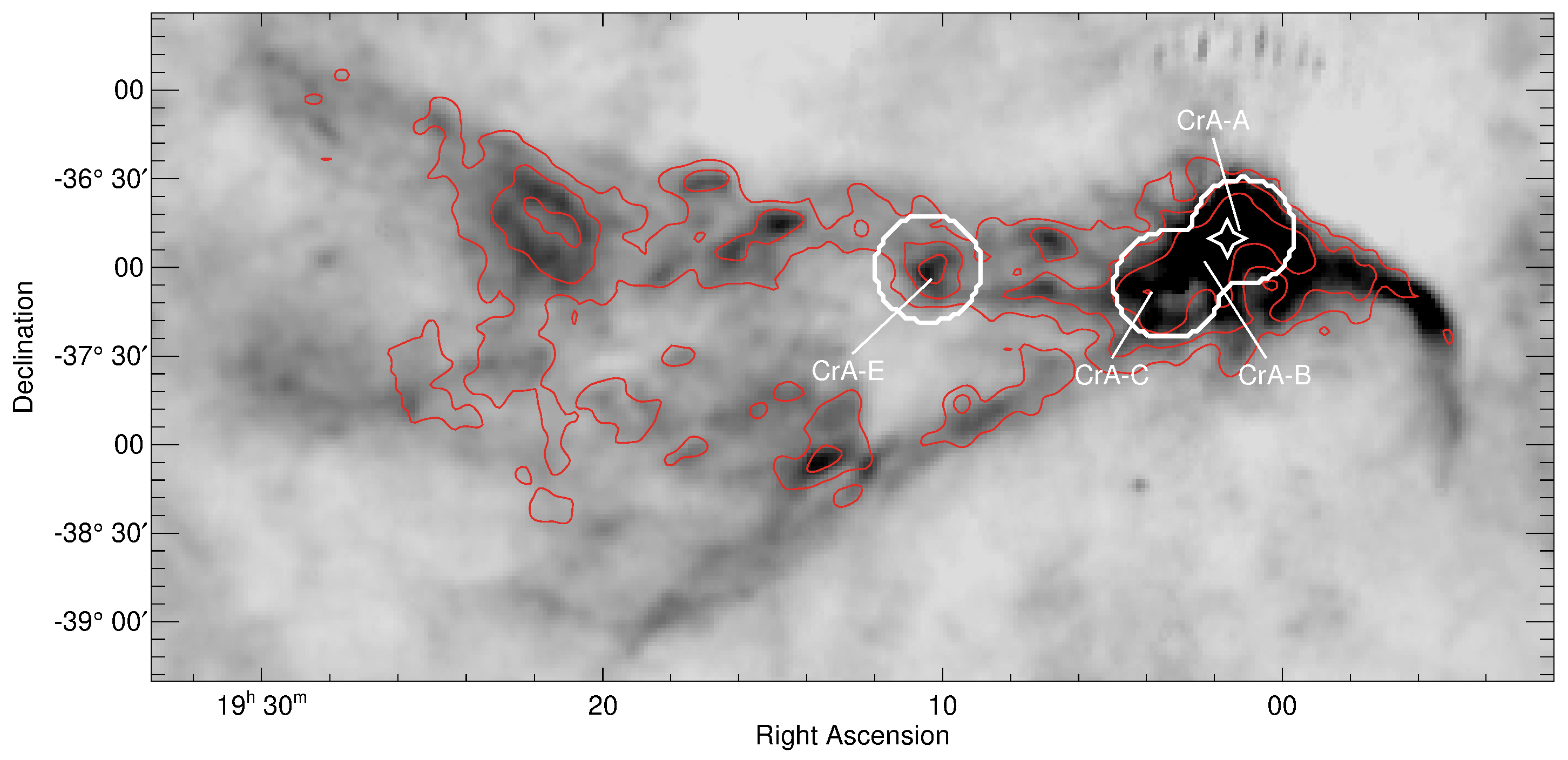}
\caption{Finding chart of the Corona Australis molecular cloud.  The background image shows \textit{IRAS} 100-\mum emission \citep{mivilledesch2005}. The overlaid red contours show $A_{\textrm{V}}$ extinctions of 0.5, 1.0, 2.0, and 5.0 magnitudes \citep{dobashi2005}. The regions observed as part of the JCMT GBLS are shown as white contours. Four subregions are labeled where data were taken. The white star towards the west of the image, in the vicinity of CrA-A, is the location of the Coronet cluster \citep{taylorandstorey1984}.}
\label{fig:findingchart}
\end{figure*}

In this paper we present SCUBA-2 observations of CrA, taken as part of the James Clerk Maxwell Telescope (JCMT) Gould Belt Legacy Survey (GBLS; \citealt{JCMTGBS2007}). Figure \ref{fig:findingchart} is a finding chart for CrA. \textit{IRAS} 100-\mum data are shown in the background \citep{mivilledesch2005}, and contours of $A_{\textrm{V}}$ are shown from \citet{dobashi2005}. The white outlines show the approximate edges of the SCUBA-2 observations (after mosaicing, see Section \ref{sec:observations}).

The region of greatest extinction is towards the west of the cloud, where the subregions CrA-A, CrA-B, and CrA-C are labeled on Figure~\ref{fig:findingchart}. This nomenclature, introduced by \citet{nutter2005} and extended by \citet{bresnahan2018}, is used throughout this paper.  These sub-regions form the nucleus of the star-forming region. Three filamentary structures can be seen in the \textit{IRAS} data. Towards the far west of the cloud, a filamentary structure, referred to as a streamer by \citet{peterson2011}, is visible. An area of low extinction (A$_{\textrm{V}} < 0.5$) can be seen to the north of this filamentary structure. To the east of the cloud nucleus, two filamentary tails are visible. The CrA-E clump, the brightest clump away from the nucleus, is found on the north tail.

CrA is kinematically associated neither with the Gould Belt nor with the Lindblad Ring \citep{neuhauser2000}. \citet{harju1993} studied the large scale structure of CrA, noting that the cloud is situated on the southern arm of the H\textsc{i} shell which is associated with the Upper Centaurus Lupus (UCL) OB association \citep{degeus1992}. The filamentary structures within the tails of CrA appear to point away from this OB association.  Recent analysis of \textit{Gaia} data has suggested that CrA is accelerating away from the Galactic Plane and from the Scorpius Centaurus OB association \citep{posch2023}, perhaps having been ejected by supernova feedback.  Alternatively, or additionally, CrA is located between two expanding H\textsc{i} shells \citep{bracco2020}, suggesting that it may be the result of a cloud-cloud collision \citep{inutsuka2015, pineda2023}.  The distance to CrA has recently been revised from $129\pm11$~pc \citep{casey98} to $151\pm9$~pc \citep{zucker2019}; in this paper we take a distance of $130$~pc for consistency with previous work \citep{bresnahan2018}.  Adopting the \citet{zucker2019} distance would increase the masses presented in this paper by a factor 1.35, increase the radii by a factor 1.16, and decrease the volume densities by a factor 0.86.  Column densities, dust temperatures and dust opacity indices would be unaffected, and none of our conclusions would be altered. 

In this paper, we identify and extract the properties of starless and protostellar cores within CrA.  This paper is laid out as follows. In Section \ref{sec:observations}, we discuss the observations, data reduction, and data processing. In Section \ref{sec:results}, we discuss the source extraction method we employed, and define the source selection and classification criteria. We discuss the observed catalogue products, including the source images. In Section \ref{sec:JCMTGBS_derivparam}, we use combinations of the SCUBA-2 data and \textit{Herschel}-derived data products to derive source parameters for our SCUBA-2-identified sources. In Section \ref{sec:jcmtgbs_discprop}, we discuss the derived properties of, and present a mass-size diagram for, our sources.  In Section~\ref{sec:beta}, we consider the evidence for a variable dust opacity index in the Coronet region.  Section~\ref{sec:summary} summarizes our results.

\section{Observations} \label{sec:observations}

\subsection{SCUBA-2}

The Corona Australis molecular cloud has been imaged with SCUBA-2 as part of the JCMT GBLS  \citep{JCMTGBS2007} under project code MJLSG35. These continuum observations were made using fully sampled 30$^{\prime}$ diameter circular regions (PONG1800 mapping mode; \citealt{bintley2014}) at three positions in CrA, at 850~$\mu$m~and 450~$\mu$m~wavelengths, with resolutions of 14\farcs1~and 9\farcs6, respectively.  The three subsections of the Corona Australis region were each observed in dry weather (JCMT Grade 2; $0.05<\tau_{225}<0.08$, where $\tau_{225}$ is the atmospheric opacity at 225\,GHz) between 2012 April 24 and 2014 August 20.  These three fields are identified in GBLS survey documentation as CrA-1 (R.A., Dec. (J2000) = 19$^{h}$01$^{m}$34$^{s}$, $-36^{\circ}$55$^{\prime}$51$^{\prime\prime}$; 6 repeats), CrA-2 (R.A., Dec. (J2000) = 19$^{h}$03$^{m}$33$^{s}$, $-$37$^{\circ}$13$^{\prime}$54$^{\prime\prime}$; 5 repeats), and CrA-E (R.A., Dec. (J2000) = 19$^{h}$10$^{m}$22$^{s}$, $-$37$^{\circ}$07$^{\prime}$49$^{\prime\prime}$; 6 repeats; GBLS nomenclature here agrees with that of \citealt{bresnahan2018})\footnote{{The detailed observation dates are as follows.  CrA-1: 2012 April 24, 2012 April 25 (twice), 2013 April 28, 2013 July 14, 2013 July 28.  CrA-2: 2013 September 29, 2014 March 15, 2014 March 18, 2014 March 19, 2014 March 20.  CrA-E: 2014 March 01, 2014 March 12, 2014 March 20, 2014 March 21, 2014 August 20 (twice).}}.  These observations collectively have the project code MJLSG35.  Two of the fields, CrA-1 and CrA-2, overlap on the bright Coronet region, with the CrA-E field located $\sim 1$\,degree to the east of the Coronet.

The data presented in this paper form part of the GBLS Data Release 2 of the SCUBA-2 data (cf. \citealt{kirk2018}). The data were reduced using the iterative map-making routine \textsl{makemap} (a subroutine of \textsc{smurf}; \citealt{chapin2013}), and then gridded to $3^{\prime\prime}$ pixels at 850~$\mu$m~and $2^{\prime\prime}$ pixels at 450~$\mu$m.  The iterations were stopped when the change in the map pixels was $<1$ per cent of the estimated map rms. The individual reduced maps were then co-added to form a mosaic of the region. A mask created using a signal-to-noise ratio cut was created for each region. The data were then re-reduced using this mask to define the areas for which the emission was significant.  

The reduction process involves the use of a $600^{\prime\prime}$ spatial filter in both the automatic masking process and the external masking reductions. Flux recovery is therefore robust for sources with Gaussian FWHM of <2.5\arcmin. Sources with FWHM between 2.5\arcmin~and 7.5\arcmin~were detected, although the sizes and fluxes of these sources were underestimated, due to Fourier components on scales larger than 5\arcmin~being removed by the filtering step. For sources larger than 7.5\arcmin, the detectability of these sources is dependent on the size of the mask used for the reduction. At our adopted distance of $130$\,pc for the Corona Australis molecular cloud, {7.5\arcmin} corresponds to $\sim0.3$ pc.  (See \citealt{mairs2015} for a detailed discussion of the role of masking in SCUBA-2 data reduction.)  {However, it should be noted that all of the sources extracted and characterized in this work have sizes significantly smaller than 2.5\arcmin, and so we can be confident that their fluxes are accurately recovered.}

The data are calibrated in mJy\,arcsec$^{-2}$ using aperture Flux Conversion Factors (FCFs) of 2340 mJy\,pW$^{-1}$\,arcsec$^{-2}$ and 4710 mJy\,pW$^{-1}$\,arcsec$^{-2}$ at 850~$\mu$m~and 450~$\mu$m, respectively, derived from average values of JCMT calibrators \citep{dempsey2013}\footnote{{We note that updated SCUBA-2 calibrations are available \citep{mairs2021}.  We retain the \citet{dempsey2013} FCFs for consistency with previous GBLS work, noting that the \citet{dempsey2013} and \citet{mairs2021} pre-2018 FCFs agree within measurement uncertainty.}}. The PONG scan pattern leads to lower noise in the map center and mosaic overlap regions, while data reduction and emission artifacts can lead to small variations in the noise over the whole map.  We found typical 1$\sigma$ noise levels of 0.07 mJy~arcsec$^{-2}$ and 2.03 mJy~arcsec$^{-2}$ for the 850$\mu$m~data and 450$\mu$m~data, respectively.  {These values were determined by performing aperture photometry on emission-free regions of the maps.}

The SCUBA-2 data presented in this paper are available, along with the masks used in the data reduction, at [DOI to be inserted in proof].

SCUBA-2 850\,$\mu$m~data can be contaminated by the $^{12}$CO $J=3\rightarrow2$ transition \citep{drabek2012,coude2016,parsons2018}, which, with a rest wavelength of 867.6~$\mu$m, is covered by the SCUBA-2 850~$\mu$m~filter bandpass (half-power bandwidth 85$\mu$m; \citealt{holland2013}).  SCUBA-2 observations can be corrected for CO contamination where $^{12}$CO $J=3\rightarrow2$ data exist \citep{sadavoy2013}, typically by using JCMT HARP-B data \citep{buckle2009}.  As HARP-B data have not been taken towards CrA, we perform a more approximate correction of a small area of the Coronet cluster using JCMT Receiver B (RxB) data ({Knee}, in prep.).  Due to the high uncertainty of this correction compared to those in other GBLS studies, and to the highly limited area over which it can be applied, we restrict discussion of CO correction to Section~\ref{sec:beta}, and present uncorrected 850$\mu$m flux densities throughout this paper.  We expect the contribution of CO emission to our 850 \um\ flux densities to be small, generally $< 20$\% except in the immediate vicinity of protostellar outflows \citep[e.g.,][]{pattle2015}.  As discussed in Section~\ref{sec:beta}, the effects of CO contamination should thus be minimal outside the centre of the Coronet cluster.

Figures \ref{fig:regmap_850} and \ref{fig:regmap_450} show the final reduced data, for regions containing significant emission. The complete data at both wavelengths are shown in Figures~\ref{fig:scubamap_850} and \ref{fig:scubamap_450} in Appendix~\ref{sec:appendix_data}. Figure \ref{fig:regmap_rgb} shows a three-colour image using \textit{Herschel} 160- and 250-\mum data (blue and green, respectively; \citealt{bresnahan2018}), and SCUBA-2 850\mum data (red).

\subsection{Herschel Space Observatory}

The ESA \textit{Herschel} Space Observatory was a 3.5-metre-diameter telescope which operated in the far-infrared and submillimetre regime  \citep{pilbratt2010}. The observations we used in this paper were taken as part of the \textit{Herschel} Gould Belt Survey (HGBS; \citealt{andre2010HGBS}). The Photodetector Array Camera and Spectrometer (PACS; \citealt{Poglitsch2010}) and the Spectral and Photometric Imaging Receiver (SPIRE; \citealt{Griffin2010}) were used in a parallel operating mode. In this mode, observations are taken simultaneously with both instruments, which scanned areas of the sky at a rate of 60\arcsec/s.  The full area observed with SCUBA-2 was covered by \textit{Herschel}. The \textit{Herschel} data we used here are presented by \citet{bresnahan2018}. We used the Herschel data for which there is a common area for SCUBA-2, PACS and SPIRE. The \textit{Herschel} data had wavelengths of 160$\mu$m, 250$\mu$m, 350$\mu$m and 500$\mu$m. The observation IDs for these data are 1342206677 -- 80, and they were reduced using the \textsl{Herschel Interactive Processing Environment} (\textsl{HIPE}; \citealt{ott2011}). The SCUBA-2 pipeline was applied to the \textit{Herschel} data to make comparisons and derive properties, as discussed in Section \ref{sec:data_process} below.

\subsection{Data processing} \label{sec:data_process}

Due to the filtering of atmospheric signal, SCUBA-2 is insensitive to structure on angular scales greater than its array size, $\sim 600$\arcsec. To make comparisons between the \textit{Herschel} data and the SCUBA-2 data, we followed the method described by \citet{sadavoy2013}.  The \textit{Herschel} data at each wavelength are added to the SCUBA-2 850\,\um\ bolometer time series, scaled to be a small perturbation on the SCUBA-2 signal.  For each \textit{Herschel} wavelength, this process results in a map that has the total flux of the combined SCUBA-2 and (scaled) \textit{Herschel} data at that wavelength. The filtered \textit{Herschel} map is then retrieved by subtracting the original SCUBA-2 reduced data from the combined SCUBA-2+\textit{Herschel} map, and reversing the applied flux scaling.

This spatial filtering removes the large-scale structure from the \textit{Herschel} maps, and also applies the SCUBA-2 mask to the \textit{Herschel} data.  One consequence of this step is that the global background levels for the \textit{Herschel} maps that are normally determined using \textit{Planck} data, are no longer needed (see, e.g., \citealt{aquila2015HGBS,marsh2016}). We applied the process to data from each \textit{Herschel} wavelength corresponding to each of the three PONG areas scanned by SCUBA-2, and then combined these spatially-filtered maps into a mosaic.

\begin{figure*}
\centering
%trim={<left> <lower> <right> <upper>}
\includegraphics[width=\textwidth]{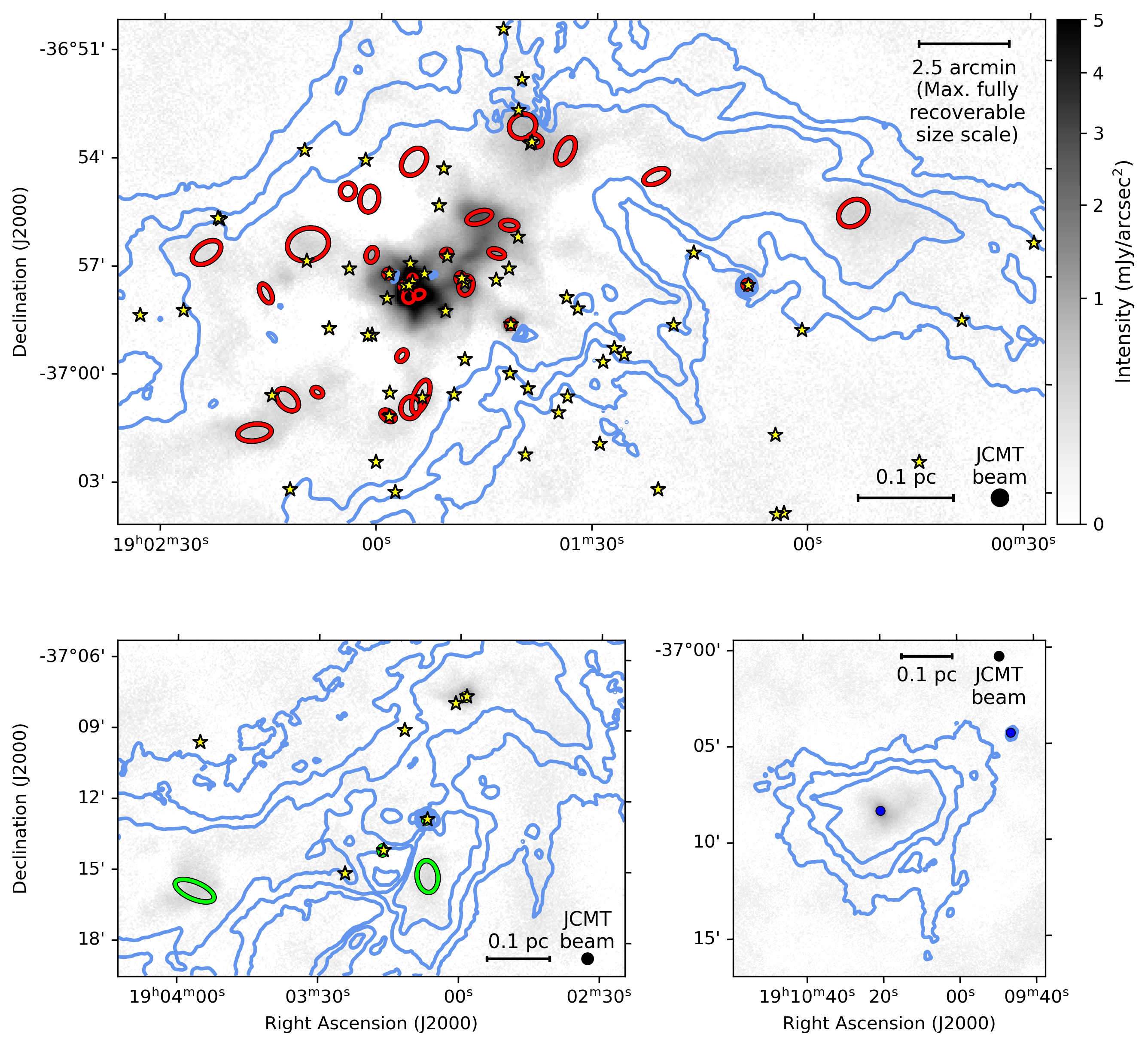}
\caption{850~$\mu$m~flux density data with a square root scaling.  Top panel: CrA-A.  Lower left panel: CrA-B and CrA-C.  Lower right panel: CrA-E.  The following colours are used for the ellipses marking our identified cores: red for CrA-A, yellow for CrA-B (one source at 19$^{h}$02$^{m}$58\dsec9 -37$^{\circ}$07$^{\prime}$37$^{\prime\prime}$), green for CrA-C and dark blue for CrA-E. The yellow stars represent YSO/protostellar candidates as found by \citet{peterson2011}, who used \textit{Spitzer} to survey Corona Australis. The contour levels are taken from the high-resolution column density map (see \citealt{bresnahan2018}), produced using \textit{Herschel} data, and start from 3 $\sigma$ and increase in levels of 1.5 times the previous level, {with values of $2.56\times10^{21}$\,cm$^{-2}$, $3.84\times10^{21}$\,cm$^{-2}$ and $5.77\times10^{21}$\,cm$^{-2}$}.}
\label{fig:regmap_850}
\end{figure*}

\begin{figure*}
\centering
%trim={<left> <lower> <right> <upper>}
\includegraphics[width=\textwidth]{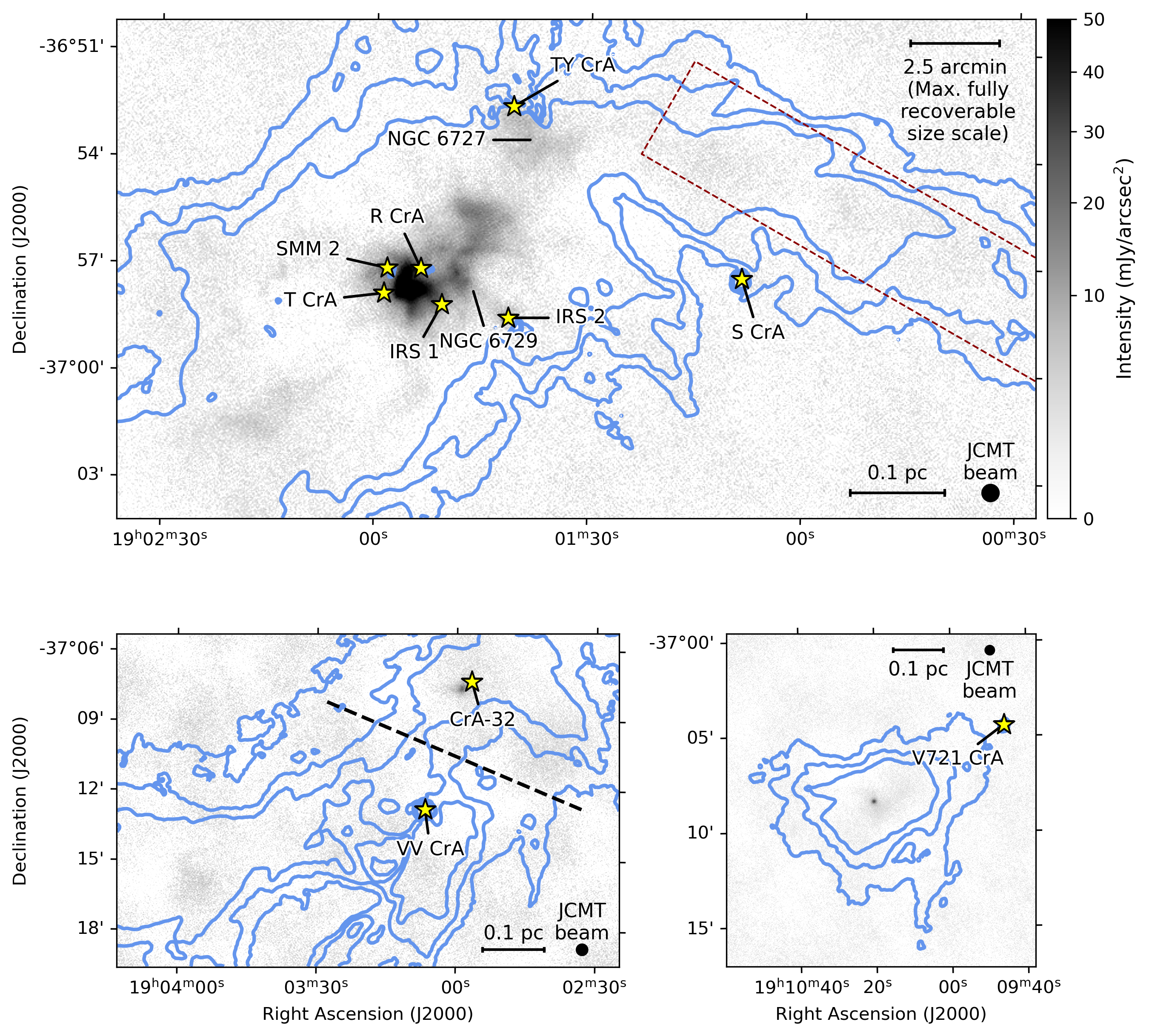}
\caption{450~$\mu$m~flux density data with a square root scaling. We show the three regions of obvious emission contained within the SCUBA-2 data. Overlaid are the contours from the high resolution column density maps from \citet{bresnahan2018}, {as in Figure~\ref{fig:regmap_850}}. The red dashed lines in the upper panel show the approximate area of the streamer/filament. The black dashed line in the lower left panel shows the approximate division line between CrA-B, which lies to the north of the line and CrA-C, which lies to the south of the line. Several young and well-studied stars are plotted on the panels as yellow stars.}
\label{fig:regmap_450}
\end{figure*}

\begin{figure*}
\centering
%trim={<left> <lower> <right> <upper>}
\includegraphics[width=\textwidth]{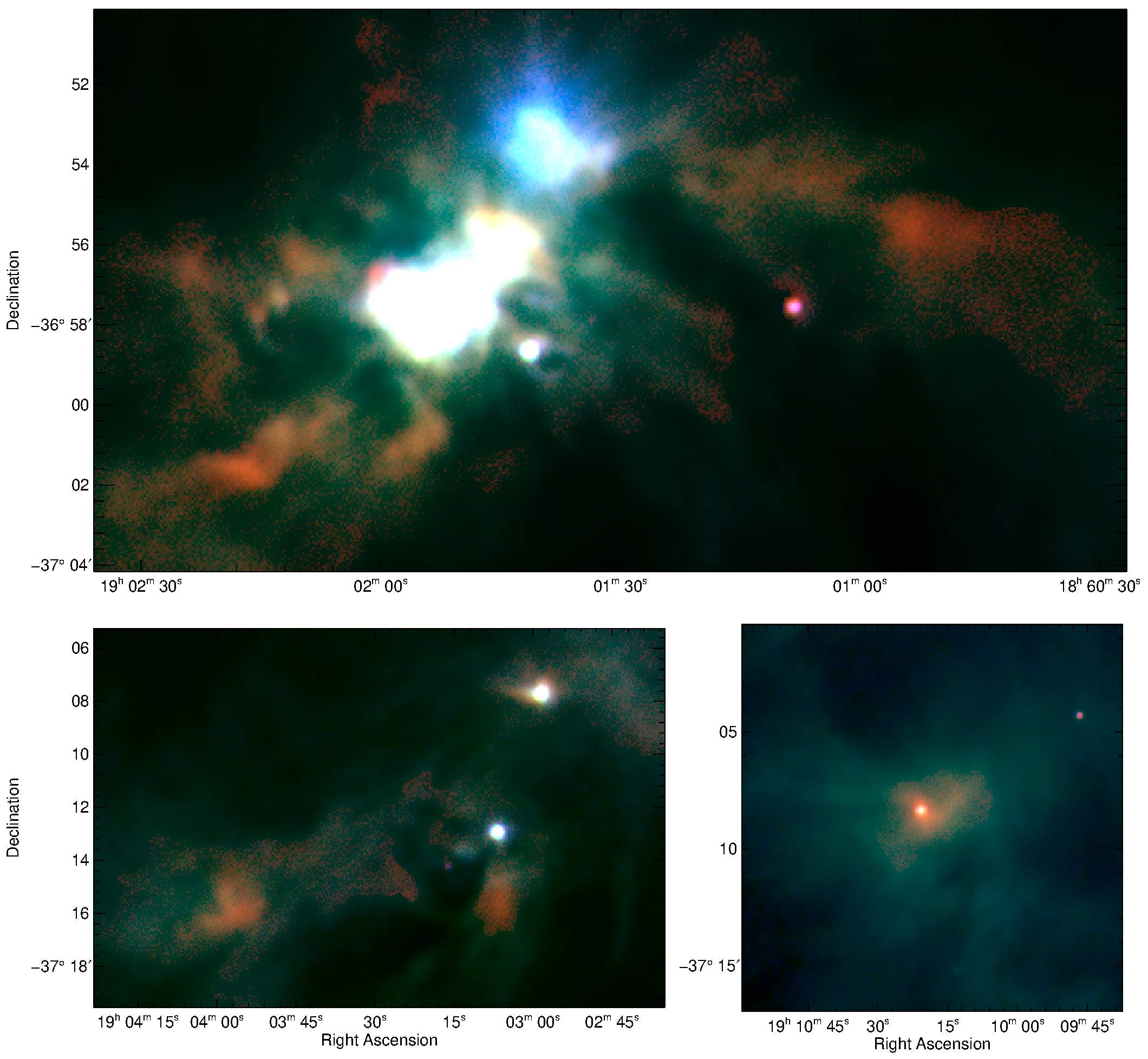}
\caption{JCMT-\textit{Herschel} three-colour images of three most visible regions within Corona Australis. Red channel: SCUBA-2 850~$\mu$m~data. Green channel: \textit{Herschel} 250-$\mu$m~data. Blue channel: \textit{Herschel} 160-$\mu$m~data.  Larger scale structure is observed by \textit{Herschel}, and the dynamic range of \textit{Herschel} images is much higher than that of SCUBA-2. This is especially prevalent in the lower-right panel of the figure. The reflection nebula NGC 6727 can be seen to the north-west of the Coronet, which is the brightest area on the upper panel.  Note that due to the significantly lower dynamic range in the SCUBA-2 data than in the \emph{Herschel} data, for aesthetic purposes we excluded 850\mum\ data at locations without significant \textit{Herschel} 250-$\mu$m emission.} 
\label{fig:regmap_rgb}
\end{figure*}

\section{Results} \label{sec:results}

The Corona Australis molecular cloud has a wide range of column densities (see \citealt{bresnahan2018}). The upper panel of Figure \ref{fig:regmap_450} shows the most active part of the star-forming region, CrA-A (see Section \ref{sec:jcmtgsselection}), which contains many of the well-studied objects in the molecular cloud. The Coronet itself contains several young stellar objects, including the well-studied Herbig Ae/Be stars R CrA and T CrA. R CrA has the spectral classification B5IIIpe \citep{gray2006}.

Adjacent to the Coronet is NGC 6729 \citep{reynolds1916}, which has shown variability in blue light measurements by \citet{graham1987}. This variability has been partially attributed to R CrA itself, which was shown to have spectroscopic variations in its Balmer line and Fe II emission profiles \citep{joy1945}. Another reflection nebula, NGC 6727, is located $\sim$5\arcmin~to the north-west of the Coronet, and is illuminated by TY CrA \citep{DWT1985CRA}.  To the west of the Coronet is the strongly accreting T Tauri star S CrA \citep{appenzeller1986,gahm2018}.

\subsection{Source extraction}
\label{sec:se}

We used the \textsl{getsources} source identification algorithm \citep{getsources2012} to locate and extract the compact and extended starless cores and protostellar objects within the data. \textsl{Getsources} is a multi-wavelength, multi-scale algorithm, produced primarily for working with \textit{Herschel} images (see, e.g., \citealt{andre2010HGBS,konyves2010,aquila2015HGBS,marsh2016}). Specifically, it was developed for extracting sources in the presence of source blending, complex backgrounds, and filamentary structure prevalent in the \textit{Herschel} GBS images. \textsl{Getsources} is also able to process data from SCUBA-2, among several other instruments from many telescopes.

We used the ``November 2013'' major release of \textsl{getsources} (1.140127) to produce our catalogue of cores. A brief description of the methodology \textsl{getsources} uses is given by \citet{aquila2015HGBS}.

For this first-generation paper from the JCMT GBLS, we used only the 850$\mu$m~data to identify sources. The dynamic range of SCUBA-2 images is not as great as in \textit{Herschel} images, due to the filtering of the large-scale structure during the reduction process. This filtering actually introduces an advantage in that the selection of a final catalogue of sources is easier.  \citet{dwt2016} found that SCUBA-2 is sensitive to the most dense cores within Taurus L1495. In effect, SCUBA-2's ability to detect low-density, unbound, starless cores, which form part of the {apparently continuous distribution of ISM structures from low-density unbound transient structures to gravitationally bound and collapsing prestellar cores} \citep{marsh2016}, is low.

\subsection{Selection and classification of reliable source detections}\label{sec:jcmtgsselection}

\begin{table*}
  \centering
  \caption{{This table lists the sources identified in the SCUBA-2 CrA observations using \emph{getsources}.  Sources are identified in the 850$\mu$m data only, and so source sizes and angles are listed alongside the 850$\mu$m flux densities.  Peak and integrated flux densities are given for every source at 850$\mu$m; 450$\mu$m values are listed for those sources with a sufficiently high SNR at that wavelength.  The regions in which the sources are located are determined following \citet{nutter2005}.}}
  \begin{tabular}{cccccccc@{\extracolsep{4pt}}ccc}
    \hline
 &  &  & \multicolumn{5}{c}{850\um}         & \multicolumn{2}{c}{450\um}   &  \\ \cline{4-8} \cline{9-10}
 & R.A. & Dec. & \multicolumn{2}{c}{FWHM size}   & Angle & \multicolumn{2}{c}{Flux Density}   & \multicolumn{2}{c}{Flux Density}   & \\ \cline{4-5} \cline{7-8} \cline{9-10} 
Index & h:m:s & $^{\circ}$:$^{\prime}$:$^{\prime\prime}$ & Major & Minor & E of N & Peak & Total & Peak & Total & Region \\
& \multicolumn{2}{c}{(J2000)}   & \multicolumn{2}{c}{arcsec}   & (deg.) & Jy/arcsec$^{2}$ & (Jy) & Jy/arcsec$^{2}$ & (Jy) & \\
\hline
1 & 19:00:54.17 & $-$36:55:18.8 & 53 & 41 & 53 & 0.081 & 1.120 & \multicolumn{2}{c}{--} & CrA-A \\
2 & 19:01:08.78 & $-$36:57:20.4 & 14 & 14 & -- & 0.562 & 0.513 & 2.920 & 2.830 & CrA-A \\
3 & 19:01:21.60 & $-$36:54:22.2 & 44 & 22 & 67 & 0.022 & 0.082 & \multicolumn{2}{c}{--} & CrA-A \\
4 & 19:01:34.23 & $-$36:53:41.6 & 49 & 27 & 27 & 0.073 & 0.524 & \multicolumn{2}{c}{--} & CrA-A \\
5 & 19:01:38.54 & $-$36:53:24.7 & 26 & 20 & 116 & 0.083 & 0.228 & \multicolumn{2}{c}{--} & CrA-A \\
6 & 19:01:40.23 & $-$36:53:00.9 & 45 & 40 & 62 & 0.126 & 1.370 & \multicolumn{2}{c}{--} & CrA-A \\
7 & 19:01:41.58 & $-$36:58:31.1 & 14 & 14 & -- & 1.310 & 1.230 & 7.930 & 7.890 & CrA-A \\
8 & 19:01:41.93 & $-$36:55:46.0 & 29 & 16 & 96 & 0.089 & 0.196 & \multicolumn{2}{c}{--} & CrA-A\\
9 & 19:01:43.62 & $-$36:56:33.3 & 27 & 14 & 106 & 0.019 & 0.082 & \multicolumn{2}{c}{--} & CrA-A \\
10 & 19:01:46.09 & $-$36:55:33.4 & 43 & 20 & 73 & 0.303 & 1.600 & \multicolumn{2}{c}{--} & CrA-A \\
11 & 19:01:47.84 & $-$36:57:27.2 & 32 & 23 & 15 & 0.270 & 1.160 & \multicolumn{2}{c}{--} & CrA-A \\
12 & 19:01:48.65 & $-$36:57:14.7 & 17 & 14 & 13 & 0.388 & 0.526 & 2.660 & 3.630 & CrA-A \\
13 & 19:01:50.59 & $-$36:56:34.3 & 17 & 14 & 81 & 0.099 & 0.086 & \multicolumn{2}{c}{--} & CrA-A \\
14 & 19:01:53.95 & $-$37:00:33.0 & 58 & 24 & 21 & 0.057 & 0.438 & \multicolumn{2}{c}{--} & CrA-A \\
15 & 19:01:54.36 & $-$36:57:43.0 & 20 & 15 & 70 & 1.090 & 2.140 & \multicolumn{2}{c}{--} & CrA-A \\
16 & 19:01:55.24 & $-$36:54:02.2 & 50 & 36 & 39 & 0.034 & 0.287 & \multicolumn{2}{c}{--} & CrA-A \\
17 & 19:01:55.27 & $-$36:57:16.8 & 14 & 14 & -- & 1.120 & 1.450 & 7.840 & 13.200 & CrA-A \\
18 & 19:01:55.35 & $-$37:00:51.4 & 37 & 32 & 19 & 0.057 & 0.399 & \multicolumn{2}{c}{--} & CrA-A \\
19 & 19:01:55.76 & $-$36:57:46.7 & 22 & 20 & 14 & 1.500 & 4.920 & \multicolumn{2}{c}{--} & CrA-A \\
20 & 19:01:56.46 & $-$36:57:29.7 & 14 & 14 & -- & 1.460 & 1.790 & 9.310 & 17.400 & CrA-A \\
21 & 19:01:56.63 & $-$36:59:25.2 & 21 & 15 & 36 & 0.035 & 0.074 & \multicolumn{2}{c}{--} & CrA-A \\
22 & 19:01:58.47 & $-$37:01:05.5 & 26 & 16 & 121 & 0.055 & 0.133 & \multicolumn{2}{c}{--} & CrA-A \\
23 & 19:01:58.67 & $-$36:57:08.9 & 14 & 14 & -- & 0.961 & 0.861 & 4.790 & 5.380 & CrA-A \\
24 & 19:02:00.99 & $-$36:56:37.9 & 25 & 18 & 17 & 0.084 & 0.157 & \multicolumn{2}{c}{--} & CrA-A  \\
25 & 19:02:01.39 & $-$36:55:05.1 & 43 & 31 & 10 & 0.046 & 0.305 & \multicolumn{2}{c}{--} & CrA-A \\
26 & 19:02:04.38 & $-$36:54:52.3 & 27 & 25 & 9 & 0.035 & 0.112 & \multicolumn{2}{c}{--} & CrA-A \\
27 & 19:02:08.34 & $-$37:00:27.4 & 19 & 14 & 123 & 0.031 & 0.047 & \multicolumn{2}{c}{--} & CrA-A \\
28 & 19:02:09.84 & $-$36:56:21.6 & 69 & 54 & 76 & 0.094 & 1.790 & \multicolumn{2}{c}{--} & CrA-A \\
29 & 19:02:12.49 & $-$37:00:40.6 & 44 & 28 & 135 & 0.076 & 0.556 & \multicolumn{2}{c}{--} & CrA-A \\
30 & 19:02:15.63 & $-$36:57:43.8 & 34 & 17 & 153 & 0.041 & 0.111 & \multicolumn{2}{c}{--} & CrA-A \\
31 & 19:02:17.02 & $-$37:01:35.7 & 56 & 28 & 83 & 0.114 & 1.080 & \multicolumn{2}{c}{--} & CrA-A \\
32 & 19:02:23.92 & $-$36:56:36.7 & 53 & 32 & 56 & 0.049 & 0.354 & \multicolumn{2}{c}{--} & CrA-A \\
33 & 19:02:58.86 & $-$37:07:37.4 & 17 & 14 & 134 & 0.604 & 1.510 & 2.650 & 7.360 & CrA-B \\
34 & 19:03:06.69 & $-$37:15:14.5 & 81 & 53 & 173 & 0.079 & 1.820 & \multicolumn{2}{c}{--} & CrA-C \\
35 & 19:03:07.02 & $-$37:12:51.1 & 14 & 14 & -- & 0.805 & 0.879 & 4.330 & 6.040 & CrA-C \\
36 & 19:03:16.32 & $-$37:14:09.1 & 23 & 17 & 178 & 0.040 & 0.066 & \multicolumn{2}{c}{--} & CrA-C \\
37 & 19:03:56.28 & $-$37:15:54.2 & 106 & 43 & 113 & 0.078 & 1.670 & \multicolumn{2}{c}{--} & CrA-C\\
38 & 19:09:46.07 & $-$37:04:26.7 & 14 & 14 & -- & 0.184 & 0.153 & 0.788 & 0.718 & CrA-E \\
39 & 19:10:20.29 & $-$37:08:26.3 & 14 & 14 & -- & 1.340 & 1.770 & 5.030 & 6.370 & CrA-E \\
\hline
  \end{tabular}
  \label{tab:sources}
\end{table*}

We introduced a scheme to eliminate spurious detections from our source catalogue. The 450$\mu$m~data from SCUBA-2 have higher noise levels than do the 850$\mu$m~data. This means that more extended objects such as starless cores are more likely to remain undetected at 450~$\mu$m~than at 850~$\mu$m. Inspection of Figure~\ref{fig:regmap_450} and comparison with Figure \ref{fig:regmap_850} shows that much of the more diffuse structure visible at 850~$\mu$m~falls below the sensitivity of SCUBA-2 at 450~$\mu$m.  As a result, we searched for sources within the 850$\mu$m~data only.

Sources that had an 850$\mu$m~monochromatic significance $\leq$7 were removed from the catalogue. This flag separates sources that are classified as reliable from those that are classified as tentative by \textsl{getsources} \citep{getsources2012}. At this stage, 51 sources were found.  {``Monochromatic significance'' is a metric used by \textit{getsources} as an analogue to peak signal-to-noise ratio (SNR).  The monochromatic significance of a source is determined by measuring signal-to-noise ratios over the multiple size scales on which \textit{getsources} makes measurements \citep{getsources2012}}. 

We then cross-matched our remaining sources with those listed in the NASA Extragalactic Database (NED; \citealt{Mazzarella2007}). We found no cross-matching NED objects within the FHWM ellipses of the 51 sources. We searched within 30\arcsec~of each source for a matching SIMBAD source. Where present, the nearest SIMBAD source is given within the electronic table (see Table \ref{table:jcmtgbs_tab_obs_cat} for a template).

We used the \textsl{CUrvature Threshold EXtractor} (CuTEx; \citealt{molinari2011}) and the \textsl{Cardiff Source-finding AlgoRithm} (\textsl{CSAR}; \citealt{kirk2013}) as alternative source finding algorithms. We cross-matched sources that were found by \textsl{CuTEx} or \textsl{CSAR} and those found by \textsl{getsources}. We checked for matching \textsl{CuTEx} and \textsl{CSAR} positions within the \textsl{getsources}-defined FWHM contour of each source, at 850~$\mu$m. Sources which have a \textsl{CSAR} or \textsl{CuTEx} match are flagged within the catalogue.

We then conducted a visual check on each source.  To do this step, we inspected the 850$\mu$m~and 450$\mu$m~data within the immediate vicinity of each nominal source. We inspected the morphology of these sources and their location within the 850$\mu$m~map.  Sources with no well-defined `core-like' morphology were discarded. The images used to perform these visual checks are included in the electronic material. Two examples are given in Figures \ref{jcmtcardexamplecore} and \ref{jcmtcardexampleproto} in Appendix~\ref{sec:appendix_gs}. Of our 51 sources, we removed twelve that were not clearly visually identifiable within the 850$\mu$m~data, leaving 39 sources in the catalogue.  The key properties of these 39 sources are listed in Table~\ref{tab:sources}: source position, observed size and position angle, and 850\,\um\ and 450\um\ flux densities.  The full \textit{getsources} output for our 39 sources is included in electronic form.  A sample of the full \textit{getsources} output is shown in Table~\ref{table:jcmtgbs_tab_obs_cat} in Appendix~\ref{sec:appendix_gs}.

We assigned our sources to the subregions given by \citet{bresnahan2018}. The nomenclature for naming the subregions was originally devised by \citet{nutter2005}, who surveyed CrA using SCUBA. Our final set of sources are shown in Figure \ref{fig:regmap_850}. We identified 32 sources within CrA-A, one source in CrA-B, four sources in CrA-C, and two sources in CrA-E. YSO and protostellar candidates catalogued by \citet{peterson2011} are also plotted.

We used results from previous surveys to categorise sources as starless cores or protostellar (and YSO) sources. We identified sources that were associated with YSOs and protostars within CrA using the \textit{Spitzer} catalogue compiled by \citet{peterson2011}, and the WISE YSO catalogue \citep{marton2015}. We checked for sources within the \textsl{getsources}-defined FWHM contour at 850~$\mu$m. Where present, the nearest matching \textit{Spitzer} or WISE source name is given in the electronic table (see Table \ref{table:jcmtgbs_tab_obs_cat}). Within the SCUBA-2 data, protostellar sources are clearly seen as bright, compact objects, on top of extended backgrounds, typically at the size of the beam. One advantage of this morphology is that protostellar sources are easily identified even in the presence of jet cavities and other surrounding material, which may be directly linked to the protostar.

We note that the source known as V721 CrA (see lower right panel of Figure \ref{fig:regmap_450}) is beyond the limit of the \textit{Spitzer} survey field.  We include this source in our catalogue as a protostar nonetheless, as it has been located in previous work and associated with the molecular cloud \citep{marraco1981,wilking1992,2masspsc2003,kohler2008,marton2015}.

Of the 39 sources we identified in CrA, {24} were found to be starless cores. The other {15} were matched to a protostellar/YSO source, and were therefore categorised as protostellar cores.  Source classifications are listed in Table~\ref{tab:classify}.

\section{SCUBA-2- and Herschel-derived parameters}\label{sec:JCMTGBS_derivparam}

\subsection{SCUBA-2-derived temperatures and masses}

We initially derived temperatures and masses for our sources using SCUBA-2 integrated flux densities at 450~$\mu$m~and 850~$\mu$m.  We convolved the 450\mum map to the resolution of the 850\mum map using the method described by \citet{aniano2011}. We used the convolution kernel derived using the same prescription as the kernels used in \citet{pattle2015} (see also \citealt{rumble2016}).

Source temperatures can be determined from the flux densities $F_{\nu_{1}}$ and $F_{\nu_{2}}$ at frequencies $\nu_1$ and $\nu_2$, respectively using the relation
\begin{equation}
\frac{F_{\nu_{1}}}{F_{\nu_{2}}} =\left( \frac{\nu_1}{\nu_2} \right)^{3+\beta} \left( \frac{\textrm{exp}(h \nu_2 / k_B T) - 1}{\textrm{exp}(h \nu_1 / k_B T) - 1} \right),
\label{tempeq}
\end{equation}
where $T$ is the source temperature and $\beta$ is the dust opacity index.

For a fixed value of $\beta$, this equation may be solved numerically for $T$, under the assumption that line-of-sight temperature variation is minimal, and that both wavelengths are tracing the same population of dust grains; c.f. \citet{shetty2009}.  We chose $\beta = 2.0$ for consistency with \textit{Herschel}-derived temperature measurements (c.f. \citealt{bresnahan2018}).

In general, the ability to derive SCUBA-2 temperatures is limited by the higher RMS noise levels in the 450$\mu$m~data (in the absence of 850\um\ flux excesses, discussed in detail below). To exclude poorly-detected sources, we only calculated SCUBA-2 temperatures for sources which have a monochromatic detection significance $\geq$7 in the 450$\mu$m~band.

Of our twelve SCUBA-2-identified protostellar cores, nine had a \textit{getsources} monochromatic detection significance $\geq$7 at 450\um, for which we derived temperatures {(sources with an 850\um\ significance $<7$ were excluded from the catalogue at an earlier stage; cf. Section~\ref{sec:se})}. A single starless core, located in CrA-E, also has a SCUBA-2-derived temperature.  These temperatures are listed in Table~\ref{tab:properties}.

Our SCUBA-2-derived temperature error estimates are given through error propagation of equation \ref{tempeq}, assuming a fixed $\beta$. Under this assumption, we derived the error of the ratio of flux densities.  The error in the ratio $R={F_{\nu_{1}}}/{F_{\nu_{2}}}$ is $\Delta R = \sqrt{\Delta F_{\nu_{1}}^{2}+ \Delta F_{\nu_{2}}^{2}}$, where $\Delta F_{\nu}=\sqrt{(f_{\nu,\textrm{cal}})^{2} + (\sigma_{\nu,\textrm{flux}}F_{\nu}^{-1})^{2}}$. The parameter $f_{\nu,\textrm{cal}}$ is the fractional calibration error at the respective SCUBA-2 wavelength (8\% and 12\% for the 450$\mu$m and 850$\mu$m~bands, respectively; \citealt{dempsey2013}). The parameter $\sigma_{\nu,\textrm{flux}}$ is the conservative \textsl{getsources}-defined measurement error, with units of mJy\,arcsec$^{-2}$.

Equation~\ref{tempeq} is insensitive to temperature in the Rayleigh-Jeans limit ($h\nu/k_{\textsc{b}T}\ll 1$).  As the 450$\mu$m~data lie on the Rayleigh-Jeans tail for temperatures $\gg32$ K, as do the 850$\mu$m~data for temperatures $\gg17$ K, the reliability of the derived temperatures for dense cores falls for SCUBA-2 at temperatures $>20$ K.  This reliability is reflected in the uncertainties associated with our higher-temperature sources.

Source masses were calculated following \citet{hildebrand1983}:
\begin{equation}
M = \frac{F_{v}^{\textrm{total}}(850~\mu \text{m})D^{2}}{\kappa_{\nu (850~\mu \text{m})} B_{\nu (850~\mu \text{m})}(T)},
\label{eqn:masseq}
\end{equation}
where $F_{\nu}(850~\mu \text{m})$ is the flux density at 850~$\mu$m, $D$ is the distance to the source, $B_{\nu (850~\mu \text{m})}(T)$ is the Planck function at temperature $T$, and $\kappa_{\nu (850~\mu \text{m})}$ is the dust mass opacity. We take $\kappa_{\nu}=0.1(\nu/1~\text{THz})^{\beta}$~cm$^{2}$g$^{-1}$ \citep{beckwith1990}, {for consistency with previous work by the JCMT GLBS \citep[e.g.,][]{pattle2015,mairs2016} and} the \textit{Herschel} GBS (e.g, \citealt{aquila2015HGBS,marsh2016}).  Where possible, the SCUBA-2-derived temperatures were used to derive a SCUBA-2-only mass.  These masses are also listed in Table~\ref{tab:properties}.

\subsection{Herschel-derived temperatures and masses} \label{sec:herschel_tm}

\begin{table}
  \centering
  \caption{{This table lists the classifications (as starless, prestellar or protostellar), and Bonnor-Ebert stability ratios, of our sources.  BE stability ratios are given for resolved sources only; sources without a matched protostar are considered prestellar if $M_{BE}/M < 2$ and starless otherwise.  Best-match protostar identifications are given for each of the protostellar sources.}}
  \begin{tabular}{cccc}
    \hline
    Index	& Type &	$M_{BE}/M$ &	Protostar ID	\\
    \hline
    1	&	Prestellar &	1.51	&	--	\\
    2	&	Protostellar & --		&	S CrA	\\
    3	&	Starless	& 15.84	&		--	\\
    4	&	Starless	& 4.09	&		--	\\
    5	&	Protostellar & 6.65	&		HD 176386A	\\
    6	&	Starless	& 2.98	&	--	\\
    7	&	Protostellar	& --	&	IRS 2	\\
    8	&	Starless	& 3.88	&		--	\\
    9	&	Starless	& 8.09	&		--	\\
    10	&	Prestellar	& 0.70	&		--	\\
    11	&	Protostellar & 1.22	&	IRS 5A	\\
    12	&	Protostellar & 0.74	&	IRS 84	\\ 
    13	&	{Protostellar}	& 4.48	&	{[GMM2009] CrA 5}	\\ 
    14	&	{Protostellar}	& 3.24	&	{[GMM2009] CrA 24}	\\ 
    15	&	Prestellar	& 0.34	&		--	\\
    16	&	Starless	& 7.13	&		--	\\
    17	&	Protostellar 	& --	&	R CrA	\\
    18	&	Starless	& 3.08	&		--	\\
    19	&	Prestellar	& 0.22	&		--	\\
    20	&	Protostellar	& --	&	IRS 7B/E	\\ 
    21	&	Starless	& 7.03	&		--	\\
    22	&	Protostellar	& 4.28	&	[GMM2009] CrA 9	\\
    23	&	Protostellar	& --	&	[GMM2009] CrA 10	\\
    24	&	Starless	& 5.86	&		--	\\
    25	&	Starless	& 5.47	&		--	\\
    26	&	Starless	& 9.60	&		--	\\
    27	&	Starless	& 7.02	&		--	\\
    28	&	Prestellar	& 1.43	&		--	\\
    29	&	Starless	& 2.28	&		--	\\
    30	&	Starless	& 7.76	&		--	\\
    31	&	Prestellar	& 1.21	&		--	\\
    32	&	Starless	& 4.83	&		--	\\
    33	&	Protostellar	& 0.24	&	IRAS 18595-3712	\\
    34	&	Prestellar	& 1.17	&		--	\\
    35	&	Protostellar	& --	&	VV CrA	\\
    36	&	{Protostellar}	& 8.88	&	{[GMM2009] CrA 34}	\\ 
    37	&	Prestellar	& 1.30	&		--	\\
    38	&	Protostellar	& --	&	V721 CrA	\\
    39	&	Prestellar	& --	&	--	\\
    \hline
  \end{tabular}
\label{tab:classify}
\end{table}

 To check the reliability of our temperature estimates, we compared our SCUBA-2-derived temperatures to the \textit{Herschel}-derived temperature map for Corona Australis (c.f. \citealt{bresnahan2018}). The \textit{Herschel} temperature map was derived by SED fitting to the 160-$\mu$m~through 500-$\mu$m~data (see, e.g., \citealt{andre2010HGBS,aquila2015HGBS}). The resolution of the \textit{Herschel}-derived temperature map is limited by the 36\arcsec~resolution of the 500-$\mu$m~data. We used our \textit{getsources}-defined non-deconvolved elliptical parameters for each source as apertures, taking the mean \textit{Herschel}-derived temperature within each aperture to be representative of the temperature of that source.  The associated \textit{Herschel} error estimates are from the error in the SED fitting process, where the \textit{Herschel}-derived temperature error map is also used in the same way to derive errors on our aperture temperatures.   These temperatures are also listed in Table~\ref{tab:properties}.

 For all 39 of our sources, we derived a hybrid `SCUBA-2-\textit{Herschel}' mass, using our \textit{Herschel}-derived temperatures and SCUBA-2 observed total 850$\mu$m~flux densities to determine a mass for each source using Equation~\ref{eqn:masseq}.  These masses are also listed in Table~\ref{tab:properties}.
  
\begin{figure}
	\begin{center}
		\begin{minipage}{1.0\linewidth}
		\resizebox{1.0\hsize}{!}{\includegraphics[angle=0]{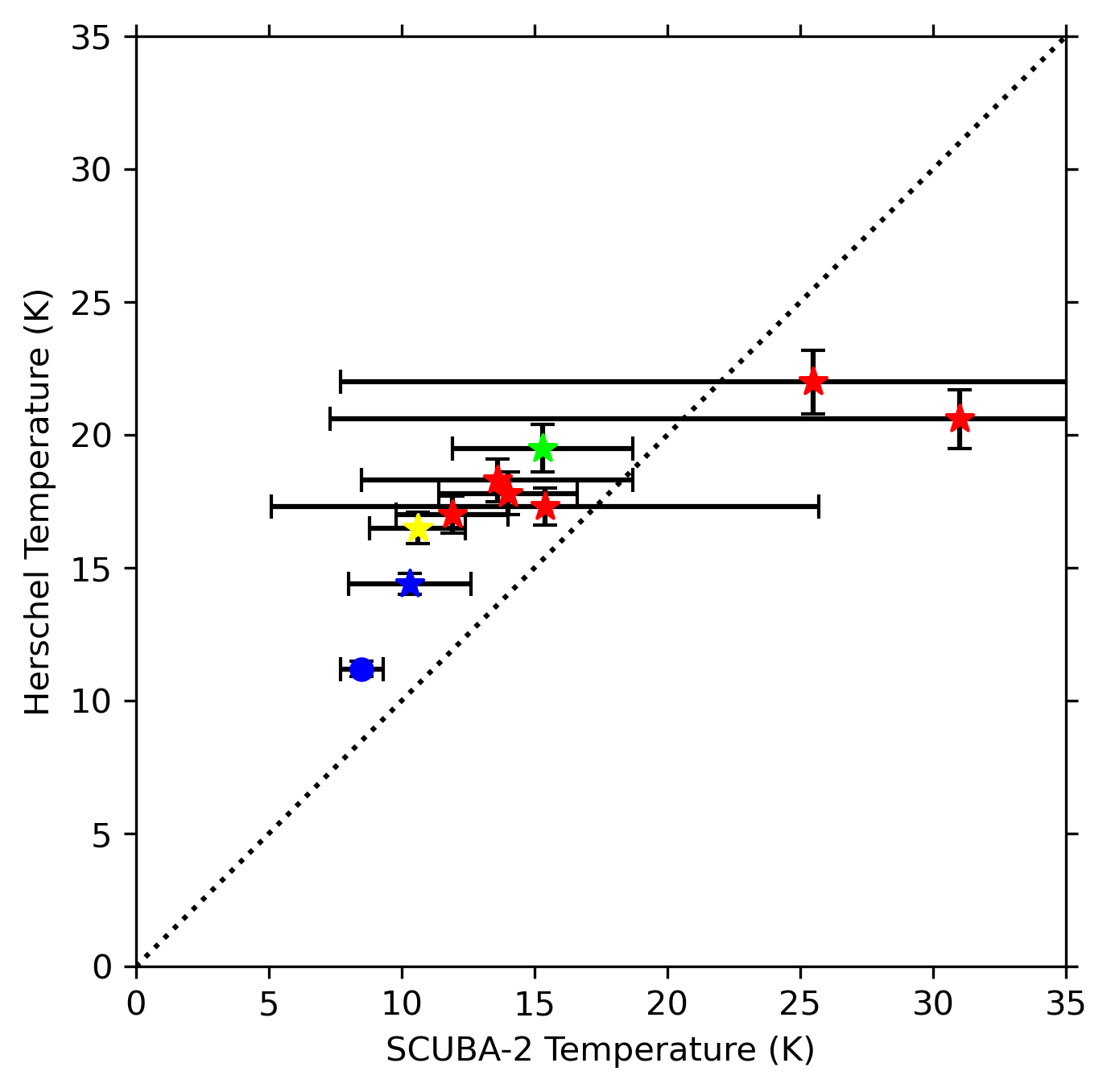}}
		\end{minipage}
	\end{center}
\caption{Comparison of the SCUBA-2-derived temperatures and \textit{Herschel}-derived temperatures, taken from the \textit{Herschel} GBS survey results. Sources are colour-coded as in Figure \ref{fig:regmap_850}. The dashed {black} line shows parity.}
     \label{s2vherschel_temps}%
\end{figure}

\begin{figure}
	\begin{center}
		\begin{minipage}{1.0\linewidth}
		\resizebox{1.0\hsize}{!}{\includegraphics[angle=0]{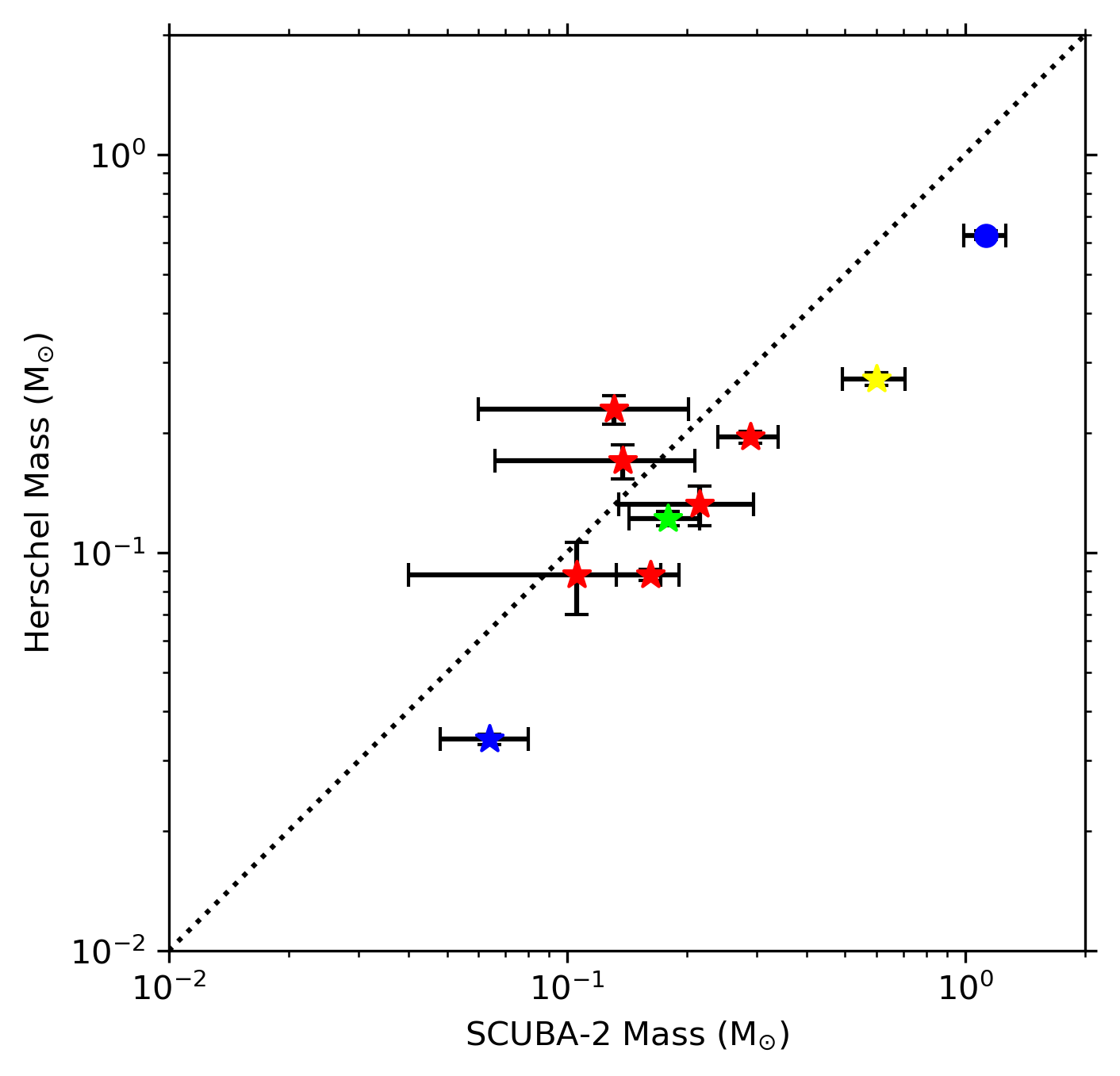}}
		\end{minipage}
	\end{center}
\caption{Comparison of our SCUBA-2-temperature-derived masses, and \textit{Herschel}-temperature-derived masses, taken from the \textit{Herschel} GBS temperature maps discussed in Section \ref{sec:herschel_tm}. Sources are colour-coded as in Figure \ref{fig:regmap_850}. The dashed {black} line shows parity.}
     \label{s2vherschel_mass}%
\end{figure}

\subsection{Comparison of derived temperatures and masses}

Figure \ref{s2vherschel_temps} shows a comparison between SCUBA-2- and \textit{Herschel}-derived temperatures. The sources are colour-coded as in Figure \ref{fig:regmap_850}. The dashed line shows where the ratio is unity. For the majority of our sources, their \textit{Herschel}-derived temperatures are higher than their SCUBA-2-derived temperatures. \citet{pattle2017} suggested that higher temperatures are usually derived from \textit{Herschel} data, due to the increased sensitivity of \textit{Herschel} observations to warmer components within the ambient cloud material. Moreover, most of our derived temperatures are for protostellar objects, where there is a greater uncertainty in source temperature (see, e.g., \citealt{pattle2017}) as determined using our SCUBA-2 data.

Figure \ref{s2vherschel_mass} shows the SCUBA-2-temperature-derived masses plotted against the \textit{Herschel}-temperature-derived masses for the ten sources with reliable SCUBA-2 temperatures. Despite the low number of sources with reliable temperatures, we find that there is no substantial disparity between our two measures of mass, with the tendency for SCUBA-2 temperatures to lead to slightly higher masses, in agreement with \citet{pattle2017}.

\subsection{Further derived properties}

In the interest of self-consistency, and given the reasonable agreement with the SCUBA-2-derived temperatures and masses, we used our \textit{Herschel}-derived temperatures and hybrid `SCUBA-2-\textit{Herschel}' masses to determine further source properties.

We derived the mean volume densities of our sources as
\begin{equation}
n( \textrm{H}_{2})= \frac{M}{\mu m_{\textrm{H}}}\frac{1}{\frac{4}{3}\pi R^{3}},
\end{equation}
where $R$ is the geometric mean of the major and minor FWHMs of the source, converted into physical distance assuming a distance of 130\,pc. Two estimates were made, for observed and deconvolved FWHM source sizes. The deconvolved source size is derived by subtracting the half-power beam-width (HPBW) of the SCUBA-2 850$\mu$m~beam (14\farcs1) from the estimated geometric mean FWHM of the source, at 850\,$\mu$m, in quadrature. We took $\mu=2.86$ following \citet{kirk2013}, assuming $\sim$70\% H$_{2}$ by mass.

We derived the mean column densities for our sources as
\begin{equation}
N(\textrm{H}_{2})= \frac{M}{\mu m_{\textrm{H}}} \frac{1}{\pi R^{2}},
\end{equation}
where the parameters are as above. Again, these quantities are given for the observed and deconvolved geometric mean FWHMs for all of our sources.  Observed and deconvolved geometric mean sizes, column densities and volume densities are listed in Table~\ref{tab:properties}.

In the absence of spectroscopic data, we used the critical Bonnor-Ebert (BE;~\citealt{bonnor1956,ebert1955}) mass, $M_{\textrm{BE}}$, of our cores to determine their dynamical states. The critical BE mass is given as 
\begin{equation}
M_{\textrm{BE,crit}} \approx 2.4 R_{\textrm{BE}}c_{\textrm{s}}(T)^{2}/G,
\end{equation}
where $R_{\textrm{BE}}$ is the BE radius, $c_{\textrm{s}}$ is the isothermal sound speed, and $G$ is the gravitational constant. The radius $R_{\textrm{BE}}$ is taken to be the deconvolved core radius measured at 850\,$\mu$m.  For this calculation, we assumed a typical gas temperature of 10\,K (in keeping with the methodology of the \emph{Herschel} GBS; we refer the reader to \citealt{aquila2015HGBS} for a justification).

Our chosen criterion for a starless core which is likely to be gravitationally bound, and so prestellar, is $M_{BE}/M < 2$, i.e. the core's derived mass is at least half of its Bonnor-Ebert mass.  This criterion is again in keeping with the \emph{Herschel} GBS (cf. \citealt{aquila2015HGBS}).  We found nine of our 27 starless cores to be bound under the critical BE criterion. We hereafter refer to these cores as prestellar cores.  The cores' $M_{BE}/M$ ratios are also listed in Table~\ref{tab:classify}.

We note that in the case of protostellar cores, the masses and temperatures determined from submillimetre dust emission are those of the protostellar envelopes, rather than those of the {embedded central hydrostatic objects or their discs}.  We further note that a given SCUBA-2 core may not necessarily have a corresponding \textit{Herschel}-identifed core, as described by \citet{bresnahan2018}\footnote{{Our cores 4, 5, 8, 9, 11, 12, 14, 16, 21, 24 and 37 do not have \citet{bresnahan2018} counterparts}}.  {The disagreements between our catalogue and \citet{bresnahan2018} mostly occur in lower-SNR regions and in the peripheries of complex structures.  We consider it likely that these differences arise from the complex response of SCUBA-2 to large-scale emission structure, and the inherent difficulty in segmenting continuous cloud structure into discrete clumps are cores.} % Why not?

\section{Discussion of properties}\label{sec:jcmtgbs_discprop}

\begin{table*}
  \setlength{\tabcolsep}{3pt}
  \centering
  \caption{{This table lists the derived properties of our sources.  Observed and deconvolved radii are listed, assuming a distance of 130\,pc and a beam size of 14.1$\arcsec$.  `S2' temperatures are given as derived from the SCUBA-2 450$\mu$m/850$\mu$m flux density ratio, for sources with sufficiently high SNR at both wavelengths.  `H' temperatures are measured from the \citet{bresnahan2018} \textit{Herschel}-derived dust temperature maps.  Masses are calculated using both temperatures.  Column densities and volume densities are calculated using `H' temperatures, for both observed and deconvolved radii.}}
  \begin{tabular}{ccc@{\extracolsep{4pt}}c@{$\pm$}cc@{$\pm$}c@{\extracolsep{4pt}}c@{$\pm$}cc@{$\pm$}c@{\extracolsep{4pt}}c@{$\pm$}cc@{$\pm$}c@{\extracolsep{4pt}}c@{$\pm$}cc@{$\pm$}c}
    \hline
     & \multicolumn{2}{c}{Radius} & \multicolumn{4}{c}{Temp.}  & \multicolumn{4}{c}{Mass}  & \multicolumn{4}{c}{$N({\rm H}_2)$} & \multicolumn{4}{c}{$n({\rm H}_{2})$} \\ \cline{2-3} \cline{4-7} \cline{8-11} \cline{12-15} \cline{16-19}
Index & (Obs.) & (Deconv.) & \multicolumn{2}{c}{S2}   & \multicolumn{2}{c}{H}   & \multicolumn{2}{c}{($T_{S2}$)}   & \multicolumn{2}{c}{($T_{H}$)}   & \multicolumn{2}{c}{(Obs.)}   & \multicolumn{2}{c}{(Deconv.)}   & \multicolumn{2}{c}{(Obs.)}   & \multicolumn{2}{c}{(Deconv.)} \\ \cline{2-3} \cline{4-7} \cline{8-11} \cline{12-15} \cline{16-19}
 & \multicolumn{2}{c}{(pc)} & \multicolumn{4}{c}{(K)} & \multicolumn{4}{c}{(M$_{\odot}$)} & \multicolumn{4}{c}{($10^{21}$ cm$^{-2}$)} & \multicolumn{4}{c}{($10^{4}$ cm$^{-3}$)} \\
\hline                                           
1 & 0.030 & 0.028 & \multicolumn{2}{c}{--} & 13.0 & 0.4 & \multicolumn{2}{c}{--} & 0.303 & 0.029 & 4.7 & 0.5 & 5.2 & 0.5 & 3.9 & 0.4 & 4.5 & 0.4 \\
2 & 0.009 & -- & 11.9 & 2.1 & 17.0 & 0.7 & 0.16 & 0.03 & 0.088 & 0.003 & 14.4 & 0.5 & \multicolumn{2}{c}{--} & 38.1 & 1.4 & \multicolumn{2}{c}{--} \\
3 & 0.020 & 0.018 & \multicolumn{2}{c}{--} & 14.5 & 0.5 & \multicolumn{2}{c}{--} & 0.018 & 0.005 & 0.6 & 0.2 & 0.8 & 0.2 & 0.8 & 0.2 & 1.1 & 0.3 \\
4 & 0.023 & 0.021 & \multicolumn{2}{c}{--} & 17.8 & 0.8 & \multicolumn{2}{c}{--} & 0.084 & 0.015 & 2.2 & 0.4 & 2.6 & 0.5 & 2.3 & 0.4 & 2.9 & 0.5 \\
5 & 0.014 & 0.011 & \multicolumn{2}{c}{--} & 21.7 & 1.2 & \multicolumn{2}{c}{--} & 0.027 & 0.003 & 1.8 & 0.2 & 3.0 & 0.4 & 3.0 & 0.4 & 6.5 & 0.8 \\
6 & 0.027 & 0.025 & \multicolumn{2}{c}{--} & 24.6 & 1.6 & \multicolumn{2}{c}{--} & 0.136 & 0.014 & 2.6 & 0.3 & 3.0 & 0.3 & 2.4 & 0.3 & 2.9 & 0.3 \\
7 & 0.009 & -- & 14.0 & 2.6 & 17.8 & 0.8 & 0.29 & 0.05 & 0.195 & 0.007 & 32.1 & 1.2 & \multicolumn{2}{c}{--} & 84.7 & 3.2 & \multicolumn{2}{c}{--} \\
8 & 0.014 & 0.010 & \multicolumn{2}{c}{--} & 14.5 & 0.5 & \multicolumn{2}{c}{--} & 0.043 & 0.017 & 3.1 & 1.2 & 5.5 & 2.1 & 5.4 & 2.1 & 12.7 & 4.9 \\
9 & 0.013 & 0.009 & \multicolumn{2}{c}{--} & 14.9 & 0.5 & \multicolumn{2}{c}{--} & 0.017 & 0.030 & 1.5 & 2.6 & 3.2 & 5.6 & 2.9 & 5.0 & 8.9 & 15.5 \\
10 & 0.019 & 0.016 & \multicolumn{2}{c}{--} & 13.9 & 0.4 & \multicolumn{2}{c}{--} & 0.379 & 0.082 & 14.9 & 3.2 & 19.6 & 4.2 & 19.2 & 4.1 & 29.0 & 6.2 \\
11 & 0.017 & 0.015 & \multicolumn{2}{c}{--} & 17.2 & 0.7 & \multicolumn{2}{c}{--} & 0.195 & 0.054 & 9.0 & 2.5 & 12.4 & 3.4 & 12.5 & 3.5 & 20.5 & 5.7 \\
12 & 0.010 & 0.004 & 15.4 & 10.3 & 17.3 & 0.7 & 0.11 & 0.07 & 0.088 & 0.018 & 12.0 & 2.5 & 75.0 & 15.7 & 29.2 & 6.1 & 452.0 & 95.0 \\
13 & 0.010 & 0.004 & \multicolumn{2}{c}{--} & 17.2 & 0.7 & \multicolumn{2}{c}{--} & 0.014 & 0.008 & 2.0 & 1.0 & 12.3 & 6.5 & 4.8 & 2.5 & 74.7 & 39.1 \\
14 & 0.024 & 0.022 & \multicolumn{2}{c}{--} & 13.5 & 0.4 & \multicolumn{2}{c}{--} & 0.110 & 0.021 & 2.7 & 0.5 & 3.1 & 0.6 & 2.7 & 0.5 & 3.5 & 0.7 \\
15 & 0.011 & 0.007 & \multicolumn{2}{c}{--} & 18.8 & 0.9 & \multicolumn{2}{c}{--} & 0.314 & 0.032 & 34.0 & 3.5 & 100.0 & 10.3 & 73.2 & 7.5 & 371.0 & 38.0 \\
16 & 0.027 & 0.025 & \multicolumn{2}{c}{--} & 15.4 & 0.5 & \multicolumn{2}{c}{--} & 0.057 & 0.015 & 1.1 & 0.3 & 1.2 & 0.3 & 1.0 & 0.3 & 1.2 & 0.3 \\
17 & 0.009 & -- & 25.5 & 17.8 & 22.0 & 1.2 & 0.14 & 0.07 & 0.170 & 0.017 & 27.9 & 2.9 & \multicolumn{2}{c}{--} & 73.6 & 7.6 & \multicolumn{2}{c}{--} \\
18 & 0.022 & 0.020 & \multicolumn{2}{c}{--} & 13.2 & 0.4 & \multicolumn{2}{c}{--} & 0.105 & 0.021 & 3.0 & 0.6 & 3.6 & 0.7 & 3.3 & 0.7 & 4.4 & 0.9 \\
19 & 0.014 & 0.010 & \multicolumn{2}{c}{--} & 18.9 & 0.9 & \multicolumn{2}{c}{--} & 0.719 & 0.051 & 54.1 & 3.9 & 99.8 & 7.1 & 96.7 & 6.9 & 242.0 & 17.2 \\
20 & 0.009 & 0.009 & 31.0 & 23.7 & 20.6 & 1.1 & 0.13 & 0.07 & 0.229 & 0.019 & 37.3 & 3.0 & 37.3 & 3.0 & 98.3 & 8.0 & 98.3 & 8.0 \\
21 & 0.012 & 0.007 & \multicolumn{2}{c}{--} & 14.5 & 0.5 & \multicolumn{2}{c}{--} & 0.016 & 0.006 & 1.7 & 0.6 & 4.4 & 1.7 & 3.5 & 1.3 & 15.0 & 5.7 \\
22 & 0.013 & 0.009 & \multicolumn{2}{c}{--} & 13.1 & 0.4 & \multicolumn{2}{c}{--} & 0.035 & 0.006 & 2.8 & 0.5 & 5.6 & 1.0 & 5.3 & 0.9 & 14.4 & 2.6 \\
23 & 0.009 & -- & 13.6 & 5.1 & 18.3 & 0.8 & 0.22 & 0.08 & 0.132 & 0.015 & 21.6 & 2.4 & \multicolumn{2}{c}{--} & 57.1 & 6.4 & \multicolumn{2}{c}{--} \\
24 & 0.014 & 0.010 & \multicolumn{2}{c}{--} & 16.4 & 0.6 & \multicolumn{2}{c}{--} & 0.029 & 0.011 & 2.1 & 0.8 & 3.7 & 1.4 & 3.6 & 1.3 & 8.5 & 3.2 \\
25 & 0.023 & 0.021 & \multicolumn{2}{c}{--} & 15.1 & 0.5 & \multicolumn{2}{c}{--} & 0.063 & 0.008 & 1.6 & 0.2 & 1.9 & 0.2 & 1.7 & 0.2 & 2.2 & 0.3 \\
26 & 0.017 & 0.014 & \multicolumn{2}{c}{--} & 15.0 & 0.5 & \multicolumn{2}{c}{--} & 0.024 & 0.002 & 1.2 & 0.1 & 1.7 & 0.2 & 1.7 & 0.2 & 2.8 & 0.3 \\
27 & 0.011 & 0.005 & \multicolumn{2}{c}{--} & 13.3 & 0.4 & \multicolumn{2}{c}{--} & 0.012 & 0.003 & 1.5 & 0.4 & 6.0 & 1.7 & 3.4 & 1.0 & 27.2 & 7.6 \\
28 & 0.039 & 0.038 & \multicolumn{2}{c}{--} & 14.0 & 0.4 & \multicolumn{2}{c}{--} & 0.425 & 0.049 & 3.9 & 0.5 & 4.2 & 0.5 & 2.5 & 0.3 & 2.7 & 0.3 \\
29 & 0.023 & 0.021 & \multicolumn{2}{c}{--} & 13.2 & 0.4 & \multicolumn{2}{c}{--} & 0.145 & 0.017 & 4.0 & 0.5 & 4.8 & 0.6 & 4.3 & 0.5 & 5.6 & 0.7 \\
30 & 0.015 & 0.012 & \multicolumn{2}{c}{--} & 14.2 & 0.4 & \multicolumn{2}{c}{--} & 0.026 & 0.006 & 1.5 & 0.3 & 2.3 & 0.5 & 2.4 & 0.5 & 4.6 & 1.0 \\
31 & 0.025 & 0.023 & \multicolumn{2}{c}{--} & 12.5 & 0.3 & \multicolumn{2}{c}{--} & 0.311 & 0.021 & 6.9 & 0.5 & 7.9 & 0.5 & 6.6 & 0.5 & 8.2 & 0.6 \\
32 & 0.026 & 0.025 & \multicolumn{2}{c}{--} & 14.1 & 0.4 & \multicolumn{2}{c}{--} & 0.082 & 0.012 & 1.7 & 0.2 & 1.9 & 0.3 & 1.5 & 0.2 & 1.8 & 0.3 \\
33 & 0.010 & 0.004 & 10.6 & 1.8 & 16.5 & 0.6 & 0.60 & 0.11 & 0.273 & 0.010 & 37.3 & 1.3 & 225.0 & 8.0 & 90.1 & 3.2 & 1336.0 & 47.6 \\
34 & 0.042 & 0.041 & \multicolumn{2}{c}{--} & 12.1 & 0.3 & \multicolumn{2}{c}{--} & 0.560 & 0.052 & 4.5 & 0.4 & 4.7 & 0.4 & 2.6 & 0.2 & 2.8 & 0.3 \\
35 & 0.009 & -- & 15.3 & 3.4 & 19.5 & 0.9 & 0.18 & 0.04 & 0.122 & 0.005 & 20.0 & 0.8 & \multicolumn{2}{c}{--} & 52.9 & 2.1 & \multicolumn{2}{c}{--} \\
36 & 0.013 & 0.008 & \multicolumn{2}{c}{--} & 14.1 & 0.4 & \multicolumn{2}{c}{--} & 0.015 & 0.003 & 1.4 & 0.3 & 3.0 & 0.5 & 2.7 & 0.5 & 8.5 & 1.6 \\
37 & 0.043 & 0.042 & \multicolumn{2}{c}{--} & 12.0 & 0.3 & \multicolumn{2}{c}{--} & 0.519 & 0.050 & 3.9 & 0.4 & 4.1 & 0.4 & 2.2 & 0.2 & 2.4 & 0.2 \\
38 & 0.009 & -- & 10.3 & 2.3 & 14.4 & 0.4 & 0.06 & 0.02 & 0.034 & 0.001 & 5.7 & 0.2 & \multicolumn{2}{c}{--} & 14.9 & 0.6 & \multicolumn{2}{c}{--} \\
39 & 0.009 & -- & 8.5 & 0.8 & 11.2 & 0.3 & 1.13 & 0.14 & 0.627 & 0.016 & 102.0 & 2.7 & \multicolumn{2}{c}{--} & 271.0 & 7.2 & \multicolumn{2}{c}{--} \\
\hline
  \end{tabular}
\label{tab:properties}
\end{table*}

\begin{figure*}
\centering
%trim={<left> <lower> <right> <upper>}
\includegraphics[width=0.8\textwidth]{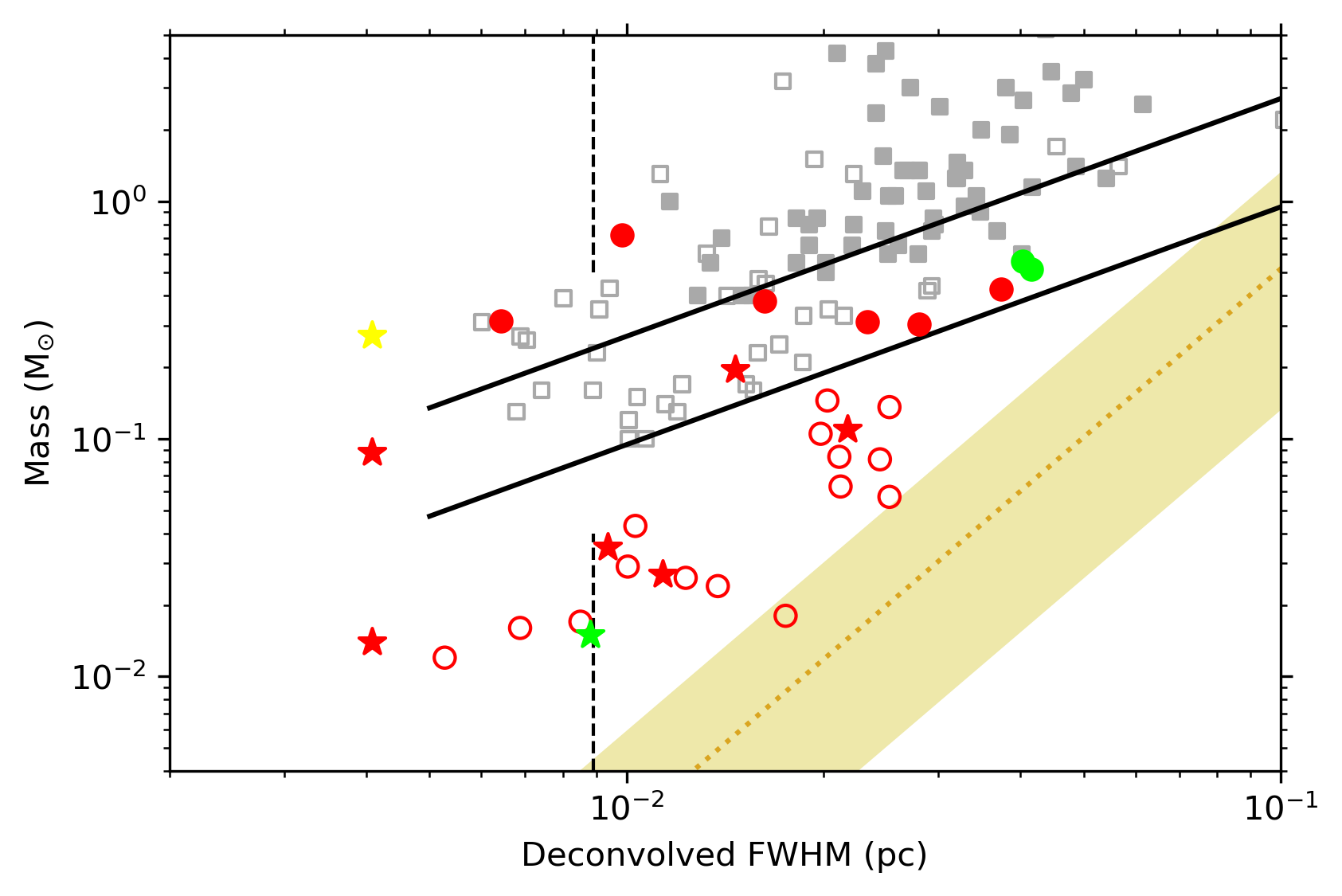}
\caption{The mass-size diagram for the population of 39 dense cores extracted by \textsl{getsources} from the SCUBA-2 data. The open circles represent cores that were classified as unbound starless cores, and are coloured by subregion. The prestellar cores are indicated by circles filled using the same colour as their respective sub-region. Protostellar cores are shown by filled stars, again, with their respective subregional colour. Shaded grey squares indicate the cores found in Orion using SCUBA \citep{motte2001}, and the open squares indicate cores found in Ophiuchus using MAMBO \citep{motte2001}. The shaded yellow band indicates the mass-size correlation observed for unbound CO clumps \citep{elemegreenfalgarone1996}. There are two model lines representing critical isothermal Bonnor-Ebert spheres, shown in black. The upper and lower lines represent BE spheres at $T = 20$ K and $T = 7$ K, respectively (c.f. \citealt{simpson2011}). The two vertical dark-grey dashed lines indicate the physical $14.1^{\prime\prime}$ resolution in the plane of sky at the assumed 130 pc distance of CrA.}
\label{fig:jcmtgbs_masssize}
\end{figure*}

Figure \ref{fig:jcmtgbs_masssize} shows the mass-size diagram for our resolved and marginally-resolved sources. On this diagram, we have split the cores into three populations according to their dynamical state: protostellar objects, prestellar cores and unbound starless cores. 

Notably, the sole prestellar core within CrA-E is unresolved by SCUBA-2. This object is marginally resolved in \textit{Herschel} observations, when extracted by \textit{getsources} \citep{bresnahan2018}. This difference is likely to be due to the fact that the extended material surrounding the core is more apparent in \textit{Herschel} images, as the dynamic range of, and sensitivity to large-scale structure in, \textit{Herschel} images is greater. This produces an inherently lower peak-to-background contrast within the \textit{Herschel} images when sources are extracted.

The two prestellar cores outside of CrA-A have larger deconvolved radii than any of the prestellar cores in CrA-A. \citet{bresnahan2018} concluded that this may be due to increased external pressures in CrA-A, created by R CrA, which acts on the surrounding dust. This external pressure acts to confine the prestellar cores in CrA-A, leading to their smaller sizes.

\begin{figure}
	\begin{center}
		\begin{minipage}{1.0\linewidth}
			\includegraphics[width=\textwidth]{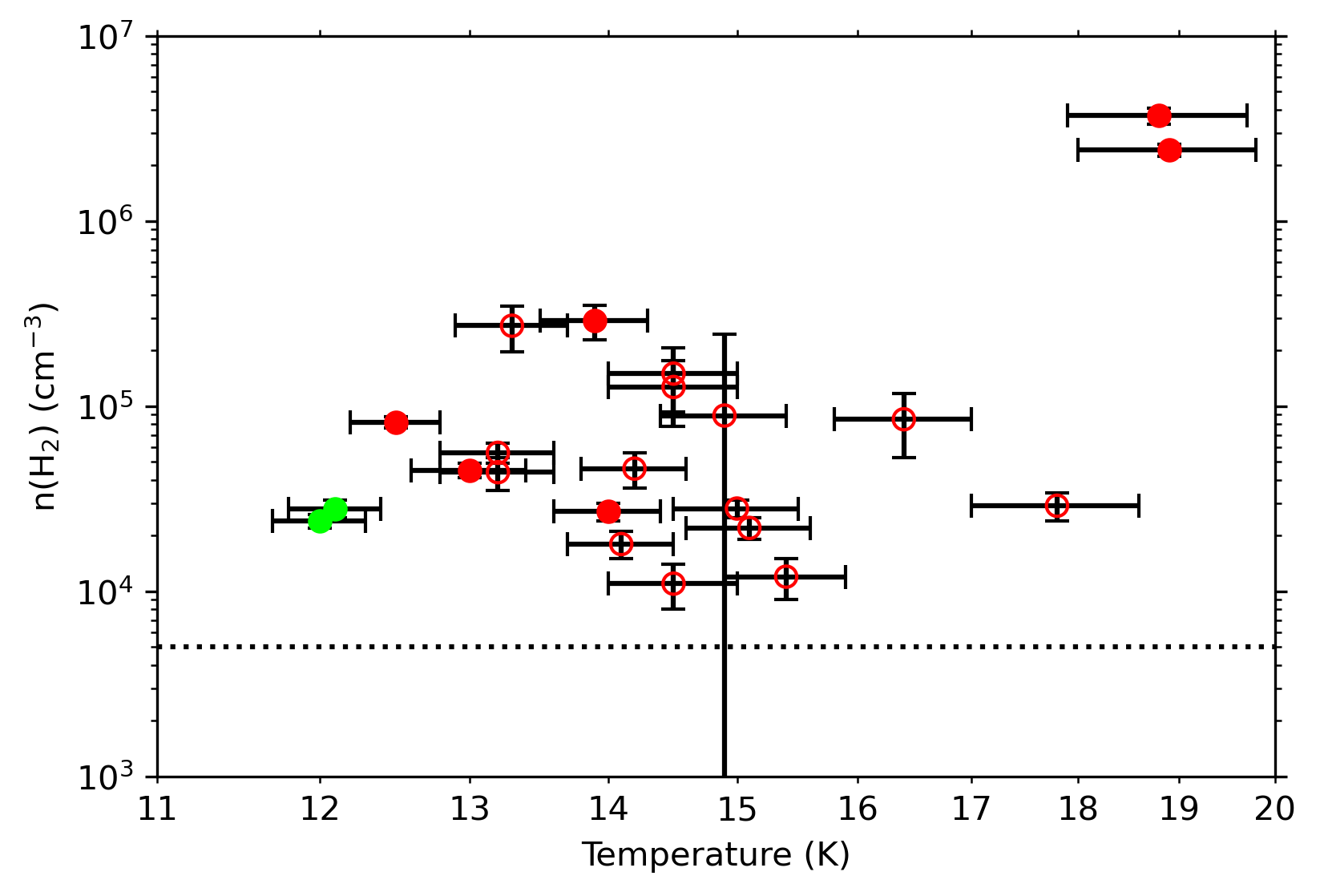}
		\end{minipage}
	\end{center}
\caption{Temperature against density of the SCUBA-2 starless cores. Sources are labelled and colour-coded as in Figure \ref{fig:jcmtgbs_masssize}. The horizontal dashed line is the $n\sim5\times10^{3}$~cm$^{-3}$ density sensitivity limit for the SCUBA-2 observations given by \citet{dwt2016}.}
     \label{fig:temp_density}%
\end{figure}

Figure \ref{fig:temp_density} shows a plot of the temperature of the starless cores against their respective densities. We do not find cores below a volume density of $\sim10^{4}$~cm$^{-3}$. \citet{dwt2016} suggested that the minimum volume density to which JCMT GBS SCUBA-2 data is sensitive for a given distance and temperature is given by
\begin{equation}
  n=n_{0}\left( \frac{D}{D_{0}} \right)^2 \frac{e^{h\nu/k_{\textrm{B}}T-1}}{e^{h\nu/k_{\textrm{B}}T_{0}-1}},
  \label{eq:empirical}
\end{equation}
where $D_{0}=140$ pc is their assumed distance to Taurus, and $T_{0}=11.3$ K is the mean density of non-externally-heated starless cores within Taurus, as determined by \citet{dwt2016}. In the case of CrA, the mean source temperature is $T=14.3$ K, and the distance is $D=130$ pc. Equation~\ref{eq:empirical} thus gives a minimum volume density sensitivity of $n\sim5\times10^{3}$~cm$^{-3}$ within CrA for the JCMT GBS. The minimum density of our starless core sample is somewhat larger than this value.  This result supports that of \citet{dwt2016} -- that there is a minimum volume density sensitivity associated with SCUBA-2 observations for a given temperature -- but suggests that more examples are required to calibrate the relation accurately.

{While temperature-density} relations have been found to follow the relation $\rho \propto T^{-a}$ in various nearby regions, using both SCUBA-2 and \textit{Herschel} data (see, e.g., \citealt{marsh2016,dwt2016,pattle2017}), {we do not observe} such a relationship for our cores.  {The reason for this lack of correlation is not clear, but may simply be the result of poor number statistics.  However, we note that the two hottest and densest of our prestellar cores, cores 15 and 19, are both located within the central Coronet region, in close proximity to the many embedded protostars, suggesting that they are subject to significant external heating.}

Comparing our range of temperatures and densities with those of \citet{marsh2016}, \citet{dwt2016}, and \citet{pattle2017}, we observe that there are a few relatively warm starless/prestellar cores within CrA-A with high densities ($> 10^{6}$\,cm$^{-3}$).  These cores are likely to be significantly heated by the nearby stars of the Coronet. 

\section{Dust opacity indices in the Coronet} \label{sec:beta}

In the earlier sections of this paper, we fixed $\beta=2$.  {However, analysis of \textit{Planck} observations gave an average $\beta$ value in nearby clouds of $1.78\pm 0.07$ \citep{planck2011}, and there is ample evidence to suggest that $\beta$ varies both within and between star-forming regions \citep[e.g.][]{juvela2015,chacon2019,tang2021}.}  Here, we investigate the consequences of determining $\beta$ through SED fitting to \textit{Herschel} PACS, \textit{Herschel} SPIRE, and JCMT SCUBA-2 wavelengths.  As discussed above, \textit{Herschel} data are sensitive to large-scale emission that is filtered out by SCUBA-2.  To compare like with like, we passed the relevant \textit{Herschel} maps through the SCUBA-2 pipeline, thus applying the same spatial filtering and masking to the \textit{Herschel} data that were applied the SCUBA-2 maps.  This technique is described in detail by \citet{sadavoy2013}.

We used the 850$\mu$m data in addition to the filtered-\textit{Herschel} data to derive temperature, column density, and $\beta$ maps towards the Coronet.  We convolved all of the filtered-\textit{Herschel} SPIRE and PACS data, as well as the SCUBA-2 850$\mu$m data, to the 36$^{\prime\prime}$ resolution of the \textit{Herschel} SPIRE 500-$\mu$m data. This convolution process is the same as that used to take the 450$\mu$m data to the resolution of the 850$\mu$m data (see \citealt{pattle2015} for more details).

The \textit{Herschel} Gould Belt Survey \citep{andre2010HGBS} derived column density maps by fitting \textit{Herschel} SPIRE and PACS bands, using non-linear least-squares fitting algorithms.  In keeping with their methodology, we used the modified blackbody equation 
\begin{equation}\label{eq:dustderiv2}
F_{\nu}=\frac{ M B_{\nu}(T_{\textrm{d}}) \kappa_{\nu}(\beta)}{D^{2}}.
\end{equation}
to fit fluxes from 160\um -- 850\um\ for the quantities $M$, $T$ and $\beta$ on a pixel-by-pixel basis.  Note that Equation~\ref{eq:dustderiv2} is functionally identical to Equation~\ref{eqn:masseq}, and that $\kappa_{\nu}(\beta)$ is defined as above.  Column density $N({\rm H_{2}}) = M/\mu m_{\textsc{h}} A$, where $A$ is the pixel area in units of cm$^{2}$ and $\mu = 2.86$, as previously. 

When performing reduced-$\chi^{2}$ fitting of a model to data, the number of free parameters that can be fitted is $n-1$, where $n$ is the number of data points to which the model is to be fitted.  However, in practice, when fitting \textit{Herschel}-only data, $\beta$ is commonly fixed, despite there typically being four wavelengths with good signal-to-noise ratios towards cold and dense regions.  The lack of longer-wavelength data ($>$500~$\mu$m) means that $\beta$, which controls the slope of the Rayleigh-Jeans tail in the modified blackbody model, generally cannot be accurately constrained \citep{sadavoy2013}.  Thus, for the \textit{Herschel} Gould Belt Survey, $\beta$ was fixed equal to 2.0 in the SED fitting process, while the temperature $T$, and dust mass $M$ were fitted to the SPIRE and PACS data.  We investigate the effect of allowing $\beta$ to vary below.

We fitted $T$, $M$, and $\beta$ to the filtered-\textit{Herschel} 160$\mu$m, 250$\mu$m, 350$\mu$m and 500$\mu$m data, and SCUBA-2 850$\mu$m data. The SCUBA-2 450$\mu$m data were excluded given the availability of the filtered-\textit{Herschel} 500$\mu$m data, which has better sensitivity.  In principle, including SCUBA-2 850\um\ data in SED fitting allows a better constraint on the value of $\beta$ by a factor of two compared to fitting \emph{Herschel} data alone \citep{sadavoy2013}.

\begin{figure*}
	\begin{center}
		\includegraphics[width=0.78\textwidth]{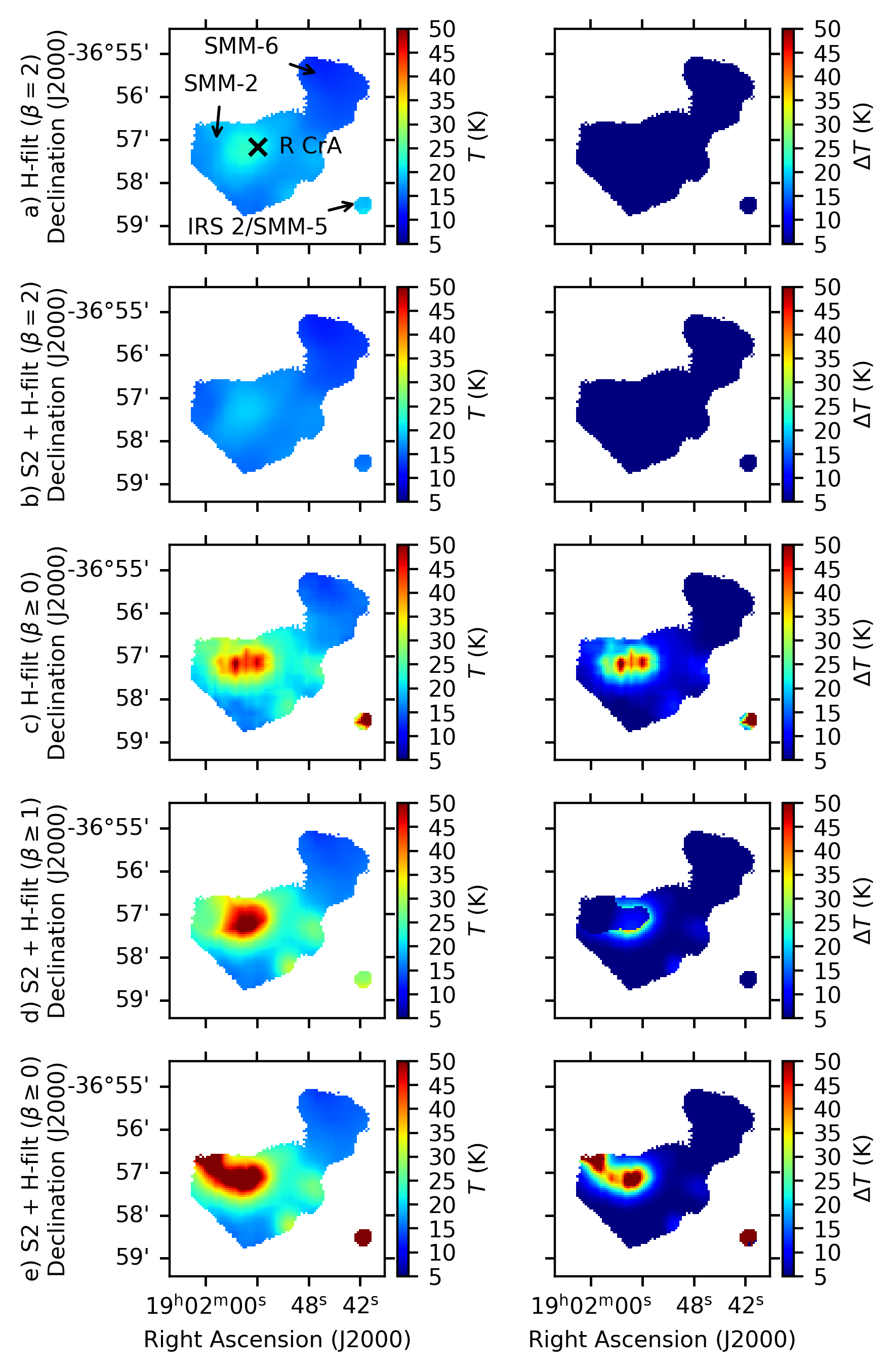}
	\end{center}
	\caption{The best-fit temperatures in the Coronet (left) and their uncertainties (right) as determined using five different sets (a)--(e) of SED fitting constraints. The cases are as follows: (a, top) filtered-\textit{Herschel} data only, with $\beta=2$; (b) filtered-\textit{Herschel} and SCUBA-2 850$\mu$m data, with $\beta=2$; (c) filtered-\textit{Herschel} only, with $\beta$ allowed to vary, but constrained to be greater than zero; (d) filtered-\textit{Herschel} and SCUBA-2 850$\mu$m data, with $\beta$ allowed to vary, but constrained to be greater than unity; and (e, bottom) filtered-\textit{Herschel} and SCUBA-2 850$\mu$m data, with $\beta$ allowed to vary, but constrained to be greater than zero.  R CrA, SMM-2, IRS 2/SMM-5 and the SMM-6 region are labelled in the upper left panel.}
	\label{fig:tempcompare}%
\end{figure*}

\begin{figure*}
	\begin{center}
		\includegraphics[width=0.78\textwidth]{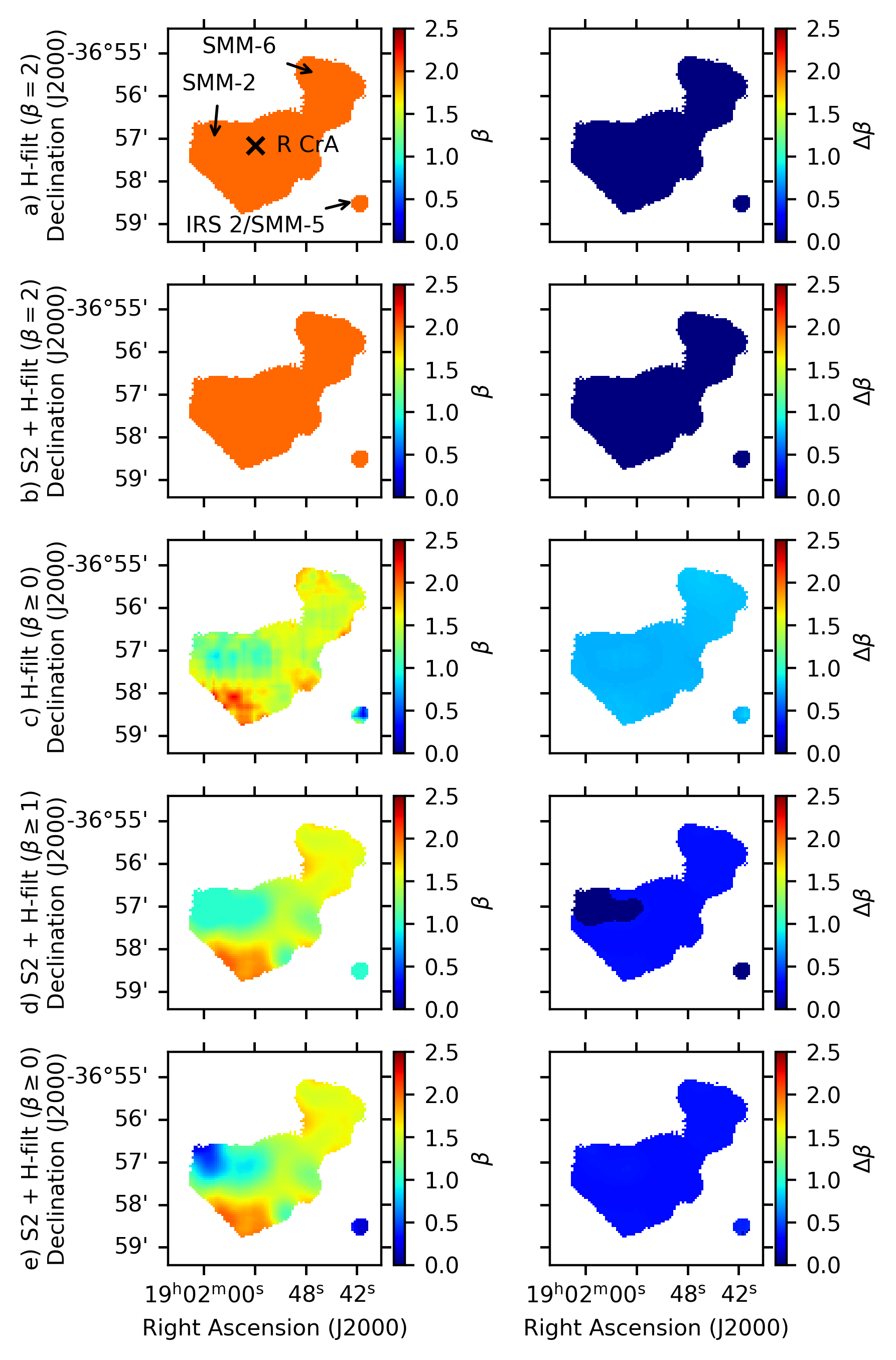}
	\end{center}
	\caption{The best-fit $\beta$ values in the Coronet (left) and their uncertainties (right). Cases (a)--(e) are the same as shown in Figure~\ref{fig:tempcompare}.}
	\label{fig:betacompare}%
\end{figure*}

\begin{figure*}
	\begin{center}
		\includegraphics[width=0.78\textwidth]{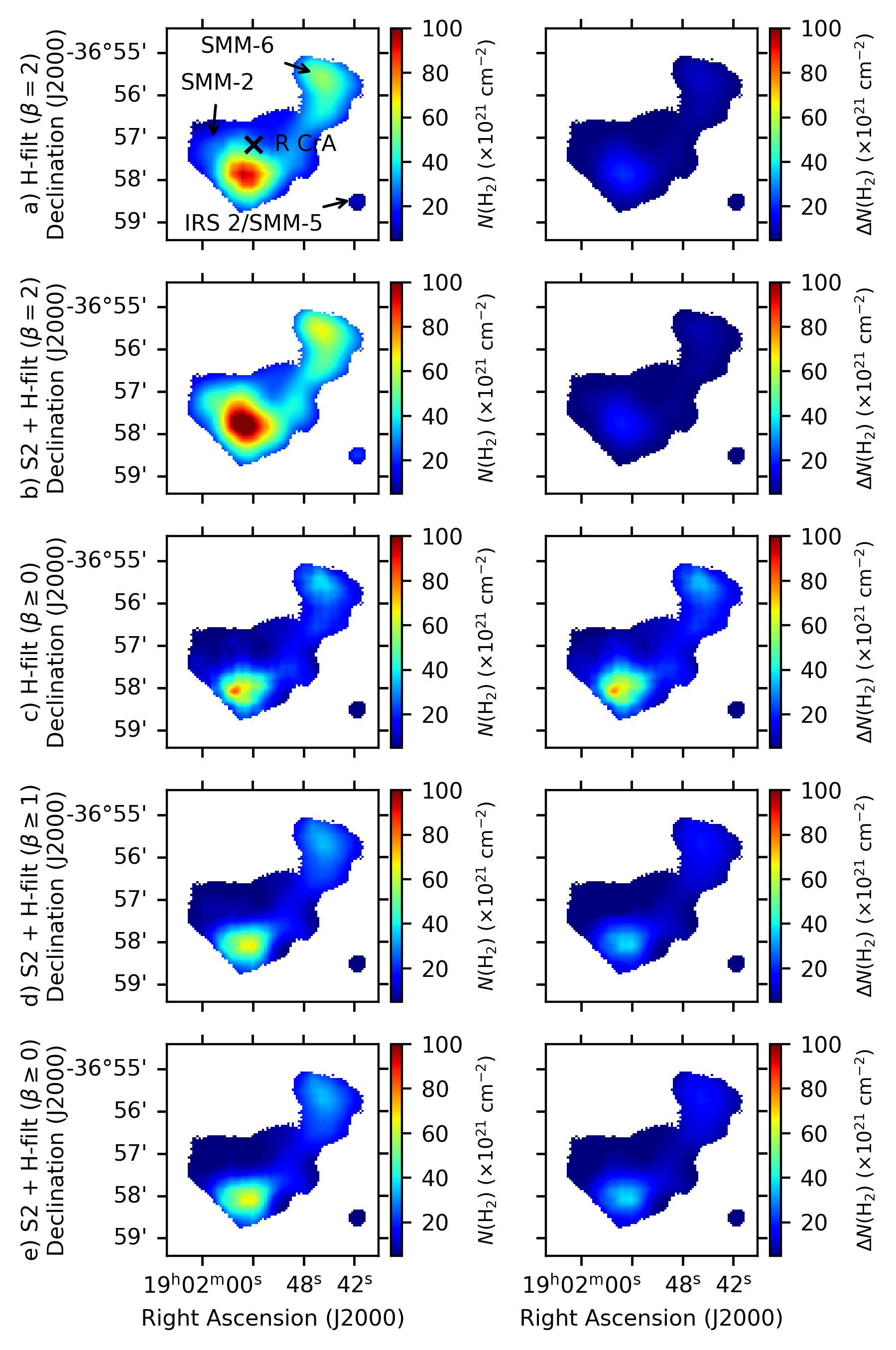}
	\end{center}
	\caption{The best-fit column densities in the Coronet (left) and their uncertainties (right). Cases (a)--(e) are the same as shown in Figure~\ref{fig:tempcompare}.}
	\label{fig:denscompare}%
\end{figure*}

Figures~\ref{fig:tempcompare}--\ref{fig:denscompare} show five combinations of wavelengths and constraints on the value of $\beta$ used to fit the emission from the Coronet region of CrA.   Figure~\ref{fig:tempcompare} shows the best-fit temperature values, Figure~\ref{fig:betacompare} shows the best-fit $\beta$ values, and Figure~\ref{fig:denscompare} shows the best-fit column density values, all with fitting uncertainties shown alongside. The cases are as follows: (a) filtered-\textit{Herschel} data only, with $\beta=2$; (b) filtered-\textit{Herschel} and SCUBA-2 850$\mu$m data, with $\beta=2$; (c) filtered-\textit{Herschel} only, with $\beta$ allowed to vary, but constrained to be greater than zero; (d) filtered-\textit{Herschel} and SCUBA-2 850$\mu$m data, with $\beta$ allowed to vary, but constrained to be greater than unity; and (e) filtered-\textit{Herschel} and SCUBA-2 850$\mu$m data with $\beta$ allowed to vary, but constrained to be greater than zero.
  
Comparing (a) and (b), including the SCUBA-2 850$\mu$m data causes the peak temperature to decrease, with the peak value of column density increasing accordingly. The morphology of the high-column-density region around the Coronet, containing the prestellar core SMM~1A, remains consistent with the \textit{Herschel}-derived column density map produced by \citet{bresnahan2018}.

Allowing $\beta$ to vary produces different behaviour in cases (c) -- (e) than in the fixed-$\beta$ cases (a) and (b).  $\beta$ is found to be $< 2$ across the Coronet, with significant variation across the region.  {Notably, it can be seen that the lowest values of $\beta$ typically correspond to the highest values of $T$, and there is not a clear (anti-)correlation between $\beta$ values and column density.  We interpret these variations in $\beta$, and correlations or lack thereof with the other fitted properties, throughout the remainder of this section.}

\subsection{Potential causes of low $\beta$ values}

Here we discuss possible causes for low values of dust opacity index $\beta$, before attempting to interpret the values of $\beta$ shown in Figure \ref{fig:betacompare}.

\subsubsection{Grain growth}

A genuinely low value of $\beta$ may be indicative of the presence of large dust grains \citep{ossenkopf1994}.  Values of $\beta$ in molecular clouds are expected to be in the range $\sim 1.5 - 2.0$ (\citealt{draine1984}; \citealt{draine2007}), while in protostellar discs $\beta \simeq 1.0$ \citep{beckwith1990}.

{Many different approaches have been taken to modelling interstellar dust, and its evolution within star-forming clouds.  These methods include numerical modelling of the properties of dust grains using tools such as the THEMIS (The Heterogeneous dust Evolution Model for Interstellar Solids) framework \citep{jones2017,ysard2019,ysard2024}, informed by laboratory analysis of interstellar dust analogues (e.g., \href{demyk2017a}{Demyk et al. 2017a}\nocite{demyk2017a},\href{demyk2017b}{b}\nocite{demyk2017b},\href{demyk2022}{2022}\nocite{demyk2022}); creating empirical dust models through fitting of observations \citep{hensley2023}; and performing hydrodynamic simulations of dust coagulation in core collapse \citep{bate2022}.  Each of these approaches demonstrates the complexity of ISM dust physics, and the sensitivity of measured dust properties, such as $\beta$, to a wide range of parameters including temperature, grain composition and UV radiation field.  However, there is a broad expectation that $\beta$ will decrease from $\sim 2$ in molecular clouds to $\sim 1$ in protostellar discs as dust coagulation occurs and maximum grain size increases \citep[e.g.][]{testi2014}, although this picture is complicated where dust grains have ice mantles \citep{ossenkopf1994}.}

{Observational results suggest significant variation of $\beta$ both within and between molecular clouds, and with Galactic environment.  \citet{juvela2015}, observing a large sample of Planck Galactic Cold Clumps (PGCCs) found a median $\beta = 1.86$, with values up to 2.2, while \citet{tang2021}, observing the Galactic Centre, found $\beta$ to increase from 2.0 to 2.4 towards dense peaks.  Values, and behaviours, of $\beta$ are also seen to vary in nearby molecular clouds and cores.  For example, \citet{bracco2017}, observing the Taurus B213 filament, found a constant $\beta = 2.4\pm0.3$ in the Miz-8b prestellar core, while finding systematic variation of $\beta$ between 1 and 2 in nearby protostellar cores.  However, \citet{chacon2019}, observing the prestellar core L1544, found $\beta$ to increase to a maximum of 1.9 in the core centre.}

Variation of $\beta$ with dust temperature has been seen in previous analyses of JCMT GBS data. \citet{chen2016}, studying the Perseus star-forming clumps, saw significant variations in $\beta$ within individual clumps.  They found that regions with low $\beta$ correlate with local temperature peaks, and demonstrated that this correlation could not be fully explained by $T-\beta$ fitting degeneracy (see Section~\ref{sec:Tbeta_degen}, below).  \citet{chen2016} argued that this effect could have resulted from grain growth in evolved clumps, {hypothesizing that grain growth} occurred while the dense clumps were cold, before the onset of star formation, but would not have resulted in a significant change in $\beta$ because the grains would also have accumulated icy mantles, which drive $\beta$ to higher values \citep{ossenkopf1994}.  Once YSOs have formed, they heat their surroundings, causing the ice to sublimate, the large grains to be exposed, and so $\beta$ to decrease.  \citet{chen2016} further suggested that protostellar outflows may be capable of carrying large, low-$\beta$ grains from deep within protostellar cores out to size scales observable by the JCMT.

{Attempts to constrain $\beta$ from far-infrared and submillimetre dust observation using SED fitting are complicated by a range of observational and instrumental effects, which we now discuss in turn.}

\subsubsection{$T-\beta$ fitting degeneracy}
\label{sec:Tbeta_degen}

Reduced-$\chi^{2}$ fitting of spectral energy distributions in the submillimetre regime {typically} results in anticorrelated values of $T$ and $\beta$ (e.g. \citealt{shetty2009}, \citealt{shetty2009a}).  The degeneracy between these two parameters can to some extent be broken by the inclusion of long-wavelength ($>$500\um) data (e.g. \citealt{sadavoy2013}). \citet{chen2016} found that fitting to both \textit{Herschel} and SCUBA-2 fluxes allows good constraints to be put on both $T$ and $\beta$ for $T<20$\,K.  Hence, in the absence of 850\,\um\ flux excesses due to CO contamination, we expect our results to be quite reliable in this temperature regime.

\subsubsection{Line-of-sight temperature variations}

Apparent $\beta$ values may be lowered by the presence of multiple dust temperature components along the line of sight.  Such temperature variation would broaden the observed SED, and hence {artificially} lower the measured value of $\beta$ {if a single-temperature model SED were fitted}.  This effect is {modelled for a two-component (10 and 15\,K) SED by \citet{shetty2009}.  \citet{malinen2011} used radiative transfer modelling of MHD simulations to show that $\beta$ can be significantly underestimated, by up to 0.5 dex, due to line-of-sight temperature varations, including in the case of cores with embedded heating sources, although they found that the effects are more severe in starless than in protostellar sources}.  While we cannot rule out such line-of-sight variations {in CrA}, we note that the spatial filtering inherent in SCUBA-2 observations (and imposed on the \emph{Herschel} observations) limits the detectability of warm, extended foreground emission.  {The} internal heating of the Coronet could create significant temperature variations along lines of sight associated with the massive protostars, {but as noted by \citet{malinen2011}, in sources with embedded protostars, the $T-\beta$ fitting degeneracy discussed in Section~\ref{sec:Tbeta_degen} is expected to have a significantly greater impact on recovered values of $\beta$ than does line-of-sight temperature variation.}

\begin{figure}
  \centering
  \includegraphics[width=0.47\textwidth]{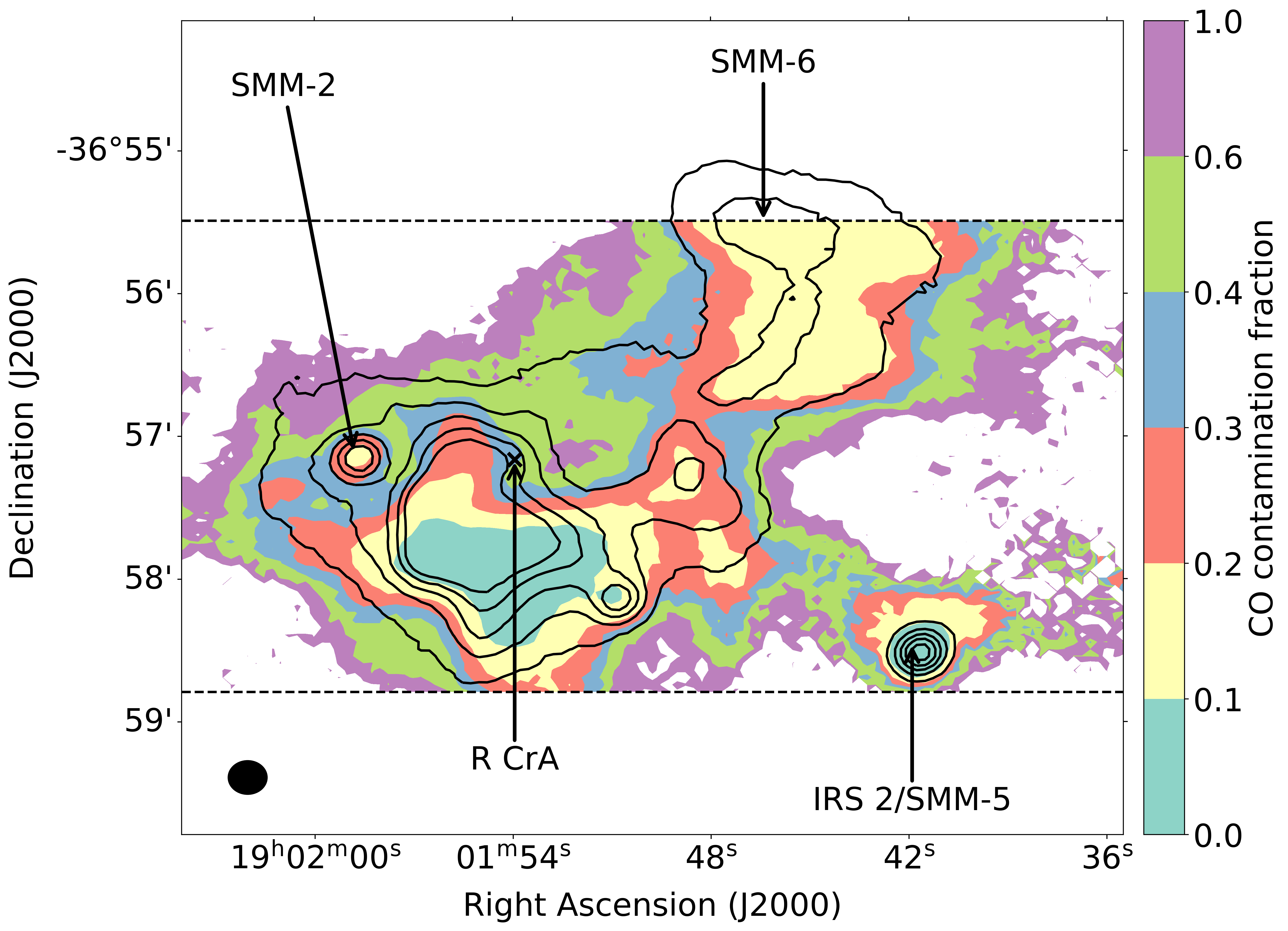}
  \caption{{Filled contours of CO contamination fraction in the Coronet cluster, with open contours of uncorrected 850~$\mu$m continuum emission overlaid.  CO contamination values are shown where 850$\mu$m intensities are greater than 0.01 mJy\,arcsec$^{-2}$.  850$\mu$m contours show 10, 20, 30, 40 and 50\% of the maximum flux density in the map.  The dashed lines show the extent of the CO map.}  Beam size is shown in the lower left-hand corner.}
  \label{fig:co_map}
\end{figure}

\subsubsection{CO contamination}
\label{sec:cofrac}

As discussed above, the SCUBA-2 850\,\um\ wide-band filter encompasses the $^{12}$CO J=$3\to2$ transition, and so 850~$\mu$m flux densities may contain an excess caused by integrated CO line emission \citep{drabek2012, holland2013, coude2016, parsons2018}.

CO contamination in JCMT Gould Belt Survey maps is typically assessed by subtracting the integrated CO $J=3\to2$ emission, as observed with HARP \citep{buckle2009}, from the SCUBA-2 timestream data (see \citealt{sadavoy2013} for a detailed description of the method).  As HARP data were not taken towards CrA, we instead tested CO contamination levels using a JCMT RxB integrated intensity map (\citealt{knee2017}; Knee, in prep.) of the Coronet region.  We added this map to the SCUBA-2 850\,\um\ bolometer time series as a negative signal and repeated the data reduction process to produce a CO-subtracted map and, by comparing this to the original map, a map of CO contamination fraction.  We find that CO contamination levels on the eastern side of the Coronet cluster, in the vicinity of R CrA and SMM-2, are unusually high, typically more than 30\%, and peaking at $>60$\%.  {However, these} values are uncertain by $\sim 20\%$ due to RxB calibration uncertainties and to the small size of the RxB map: SCUBA-2-pipeline background subtraction requires a large region of minimal astrophysical flux for good accuracy \citep{chapin2013}.  {As we discuss in Section \ref{sec:beta_smm6} below, comparison of our \textit{Herschel}-only and \textit{Herschel}+SCUBA-2 fitting results suggest that the CO contamination fractions that we derive are likely to be systematically overestimated.}   

{A filled contour plot of CO contamination fraction is shown with contours of uncorrected 850~$\mu$m continuum emission overlaid} in Figure~\ref{fig:co_map}.  Note that CO contamination fractions measured in regions of low 850\,\um\ brightness are not physically meaningful.  While the CO contamination values shown are highly uncertain, they provide useful, if somewhat qualitative, information on the location of CO-contamination-induced 850\,\um\ flux excesses in the Coronet, as discussed below.

\subsubsection{Free-free contamination}

In the vicinity of high-mass stars, free-free emission can be sufficiently strong to significantly contaminate the 850\um\ channel, contributing up to $\gtrsim 10$\% of the total flux density \citep{rumble2016}.  We do not have any evidence by which to judge levels of free-free contamination in the Coronet.  However, as the region contains embedded young massive stars, free-free emission is a possible contributing factor to 850\um\ flux excesses, particularly in the immediate vicinity of those stars. 

\subsubsection{Effect of contamination of 850$\mu$m band}
\label{sec:mc_contam}

{We used Monte Carlo methods to model the effect that adding excess flux to the 850$\mu$m data point has on the results of SED fitting, in order to quantify the potential impact of CO or free-free contamination on our results.} 

{We created modified blackbody distributions using the \citet{beckwith1990} dust opacity law used in our mass determinations above.  We chose a column density of $5\times 10^{21}$\,cm$^{-2}$ and temperatures of 10, 15 and 20\,K, and varied $\beta$ in the range 1.0--2.2.  We then added excess flux to the 850$\mu$m data point, to model the effect of contamination, testing contamination fractions in the range 0--50\%.  We next added noise to each data point drawn from a Gaussian distribution with a 1-$\sigma$ width of 10\% of the flux at that wavelength\footnote{{We also tested a 1-$\sigma$ width of 5\% of the flux, which produced almost identical results to the 10\% case.}}.  Finally, we fitted equation~\ref{eq:dustderiv2} to our model data using the \textsc{scipy} \textit{curve\_fit} routine.  We repeated this process 10\,000 times for each contamination fraction, and examined the returned $\beta$ and $T$ values.}

\begin{figure}
    \centering
    \includegraphics[width=\linewidth]{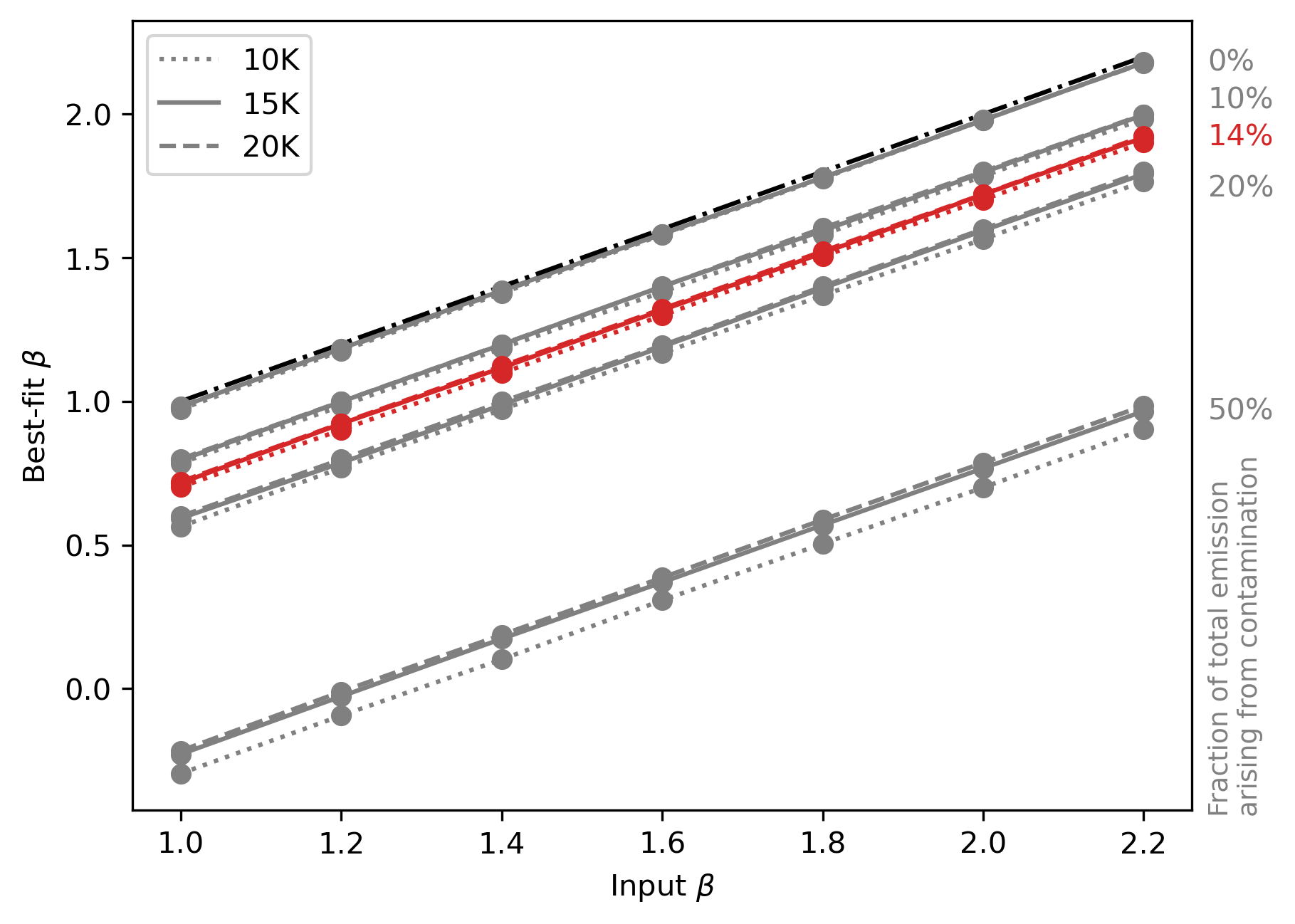}
    \includegraphics[width=\linewidth]{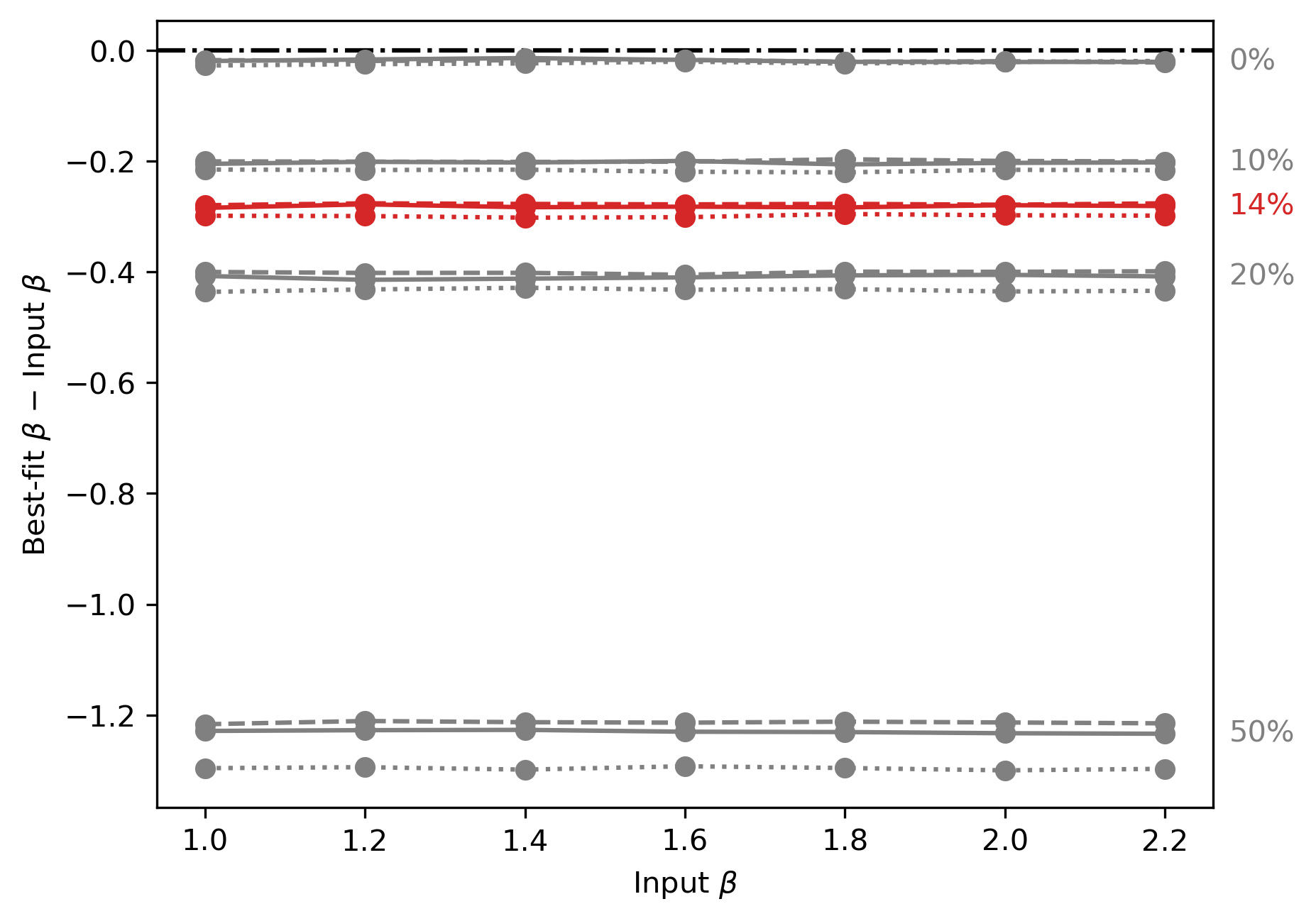}    
    \caption{{In this figure, we show the results of our Monte Carlo modelling of the effects of CO/free-free contamination on SED fitting.  Top panel: input $\beta$ vs. returned $\beta$.  Bottom panel: input $\beta$ vs. the difference between best-fit and input $\beta$.  The case in which 14\% of the total emission arises from contamination is shown in red, as this is the measured contamination in the SMM-6 region.}}
    \label{fig:co_model}
\end{figure}

{The results of this analysis are shown in Figure~\ref{fig:co_model}.  It can immediately be seen that even 10\% 850$\mu$m contamination can significantly alter the returned value of $\beta$.  Therefore, if the CO contamination fractions presented in Section~\ref{sec:cofrac} are accurate, our \textit{Herschel}+SCUBA-2 $\beta$ values are likely to be significantly underestimated, and should moreover be significantly lower than the $\beta$ values determined from the \textit{Herschel}-only fitting at the same position.  However, as discussed in Section~\ref{sec:beta_smm6}, below, comparison of our \textit{Herschel}+SCUBA-2 and \textit{Herschel}-only $\beta$ values does not provide evidence for significant differences between the two values, suggesting that our CO contamination fractions may be overestimated.}

\subsection{The R CrA/SMM-2 region}
\label{sec:beta_rcra}

When $\beta$ is allowed to vary, it shows {low} values in the vicinity of R CrA, and particularly towards the submm-bright point source SMM-2 \citep{nutter2005}, identified as a Class I protostar \citep{sandell2021}, whether or not SCUBA-2 850\um\ data are used in the fitting process.  {In the following discussion, the R CrA region is defined as R.A. $> 19^{h}01^{m}52$\fs8, Dec. $>-36^{\circ}57^{\prime}30^{\prime\prime}$, and where 850\,$\mu$m flux density is $> 2$ mJy/arcsec$^{2}$ (this latter criterion is chosen for consistency with our definition of SMM-6, below).  The SMM-2 region refers to the circular area around the SMM-2 protostar of radius 14.1 arcsec.}

When fitted using Herschel data alone, {the R CrA region as a whole} is found to be the hottest region of the Coronet, {with a median temperature of 37\,K and a median fitting error of 27\,K, as seen in Panel (c)} of Figure~\ref{fig:tempcompare}. {Panel (c) of Figure~\ref{fig:betacompare} shows a corresponding decrease in $\beta$, with a median value of 1.16 and a median uncertainty of 0.75.}  {At the position of the SMM-2 protostar, we measure a median temperature of $31\pm18$\,K, and a median $\beta$ of $1.07\pm 0.75$.}  {We might expect to see high temperatures in a region containing an embedded B5 star (R CrA), and so these low values of $\beta$, and their large uncertainties, are likely to result from $T-\beta$ fitting degeneracy rather than providing evidence for grain growth.}

The artificially low values of $\beta$ seen in the R CrA/SMM-2 region are exacerbated by the inclusion of 850\um\ data, as seen in Panels (d) and (e) of {Figure~\ref{fig:betacompare}}.  {The median temperature across R CrA becomes $48\pm30$\,K, while the median $\beta$ becomes $0.93\pm 0.35$, while the median values in the vicinity of SMM-2 become $T=48\pm 31$\,K and $\beta=0.61\pm 0.35$.}  {Figure~\ref{fig:co_map} shows that the highest CO fractions in the Coronet region occur between R CrA and SMM-2. The average CO fraction across the R CrA region as a whole is $34\%\pm10\%$, with a maximum value of 60\%.  The average CO fraction around SMM-2 is $32\%\pm10\%$, with a minimum of 10\% at the position of SMM-2 itself.}  Thus, {while our analysis of the SMM-6 region, below, suggests that these CO contamination fractions may be overestimated,} there is a strong probability that the extremely low $\beta$ values inferred in this region using JCMT 850\um\ data are artificial.

\subsection{The IRS 2 protostar}
\label{sec:beta_irs2}

The environment of the IRS 2 protostar (Class I; SMM-5 in the nomenclature of \citealt{nutter2005}), detached from the main Coronet region in the south-west of Figure~\ref{fig:tempcompare} and \ref{fig:denscompare}, presents an interesting puzzle.  {In \emph{Herschel}-only fitting, $\beta = 0.89\pm 0.76$ across this source, with $\beta$ lower on the western side.  In SCUBA-2+\emph{Herschel} fitting, $\beta = 0.26\pm 0.38$, with $T$ artificially high ($38\pm29$\,K with \textit{Herschel} only; $107\pm184$\,K with the SCUBA-2 data point)} and $N({\rm H_{2}})$ artificially low.  This source appears to have a significant long-wavelength flux excess.  As can be seen in Figure~\ref{fig:co_map}, {the CO contamination associated with IRS 2 is low, at $9\pm 6$\%, and as discussed below, this may be an overestimate}.  {These temperature and $\beta$ values are median fitted values and median fitting error measured over pixels within a radius of one JCMT beam of the source position.}

IRS 2 is a low-mass protostar, with associated X-ray emission \citep{forbrich2007}.  This system also has an associated bright rim \citep{siciliaaguilar2013}, suggesting that it may be undergoing shock heating.  If this is in fact the case, it might suggest that the artificially low values of $\beta$ seen are the result of a severe $T-\beta$ fitting degeneracy.  Another possibility is a genuine flux excess at long wavelengths, caused by large amounts of cold dust associated with the protostar.  We do not have sufficient information to distinguish between these hypotheses, but note this as an interesting source for further study.

\begin{figure*}
  \centering
  \includegraphics[width=0.8\textwidth]{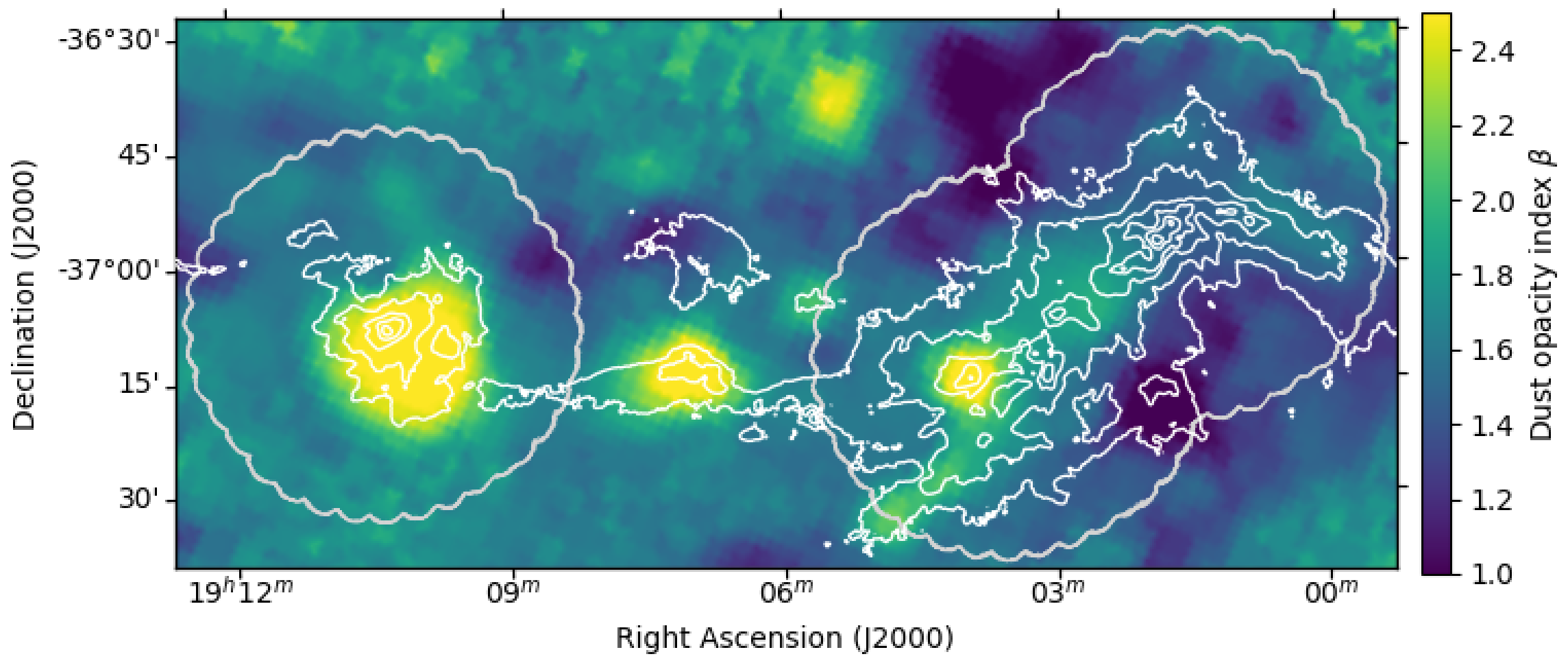}
  \caption{\textit{Planck} $\beta$ map of CrA.  White contours are \textit{Herschel} column density.  Grey outline shows regions observed {using} SCUBA-2.}
  \label{fig:planck}
\end{figure*}

\subsection{The SMM-6 region}
\label{sec:beta_smm6}

The SMM-6 region (\citealt{nutter2005}), in the north-west of the Coronet, shows a stable and plausible $\beta$ value, {$\beta\approx 1.54$, both with and without SCUBA-2 data: for \textit{Herschel}-only fitting, $\beta = 1.53\pm 0.79$, while when the SCUBA-2 data point is included, $\beta=1.55\pm 0.35$.}  This region contains no embedded protostars, and so no significant internal temperature variations are expected.  The {best-fit temperature of the clump is $15.8\pm4.2$\,K (\textit{Herschel}-only) or $15.3\pm2.3$\,K (with SCUBA-2), suggesting that $T-\beta$ degeneracy effects on the derived value of $\beta$ should be minimal.  We define SMM-6 as having R.A. $< 19^{h}01^{m}$49\fs5, Dec. $>-36^{\circ}56^{\prime}51^{\prime\prime}$, and where 850\,$\mu$m flux density is $> 2$ mJy/arcsec$^{2}$, the intensity contour at which SMM-6 separates from the rest of the Coronet region.}

{The CO contamination fraction in SMM-6 is relatively low, with a mean value of $14\pm3$\%.}
{Despite being relatively low, this CO fraction is large enough that we would expect based on our Monte Carlo modelling to see a lower $\beta$ value for the \textit{Herschel}+SCUBA-2 fit than for the \textit{Herschel}-only fit.}
{Based on our modelling described in Section~\ref{sec:mc_contam} above, a CO contamination fraction of 14\% produces a value of $\beta$ offset by $-0.28$ from the true value at a temperature of 15\,K.  We thus might expect that if we measure $\beta=1.55$, the underlying true $\beta$ would be approximately 1.83.  However, the \textit{Herschel}-only fitting produces a value almost identical to the \textit{Herschel}+SCUBA-2 fit.  The fact that we see such similar values with and without the SCUBA-2 data point included in the fit implies that the RxB CO contamination fractions given in Section~\ref{sec:cofrac} may be systematically overestimated.  We noted in Section~\ref{sec:cofrac} that there is a $\sim20$\% systematic uncertainty on RxB-derived CO contamination values; our results in SMM-6 suggest that the systematic error on our measured CO fractions is indeed large.}

{The consistent, well-constrained and plausible values of both $T$ and $\beta$ that we obtain in SMM-6 suggest that this region is} our best candidate for demonstrating grain growth within the Coronet.  {We therefore next consider the timescale for grain coagulation in molecular clouds, $t_{coag}$, following \citet{chakrabarti2005}:}
\begin{equation}
    t_{coag} \sim 6\times10^{6}\left(\frac{10^{5}\,{\rm cm}^{-3}}{n_{\rm H}} \right)\left(\frac{a}{0.1\,\mu{\rm m}}\right)\left(\frac{1\,{\rm m\,s}^{-1}}{v_{rel}}\right)\,{\rm yr},
    \label{eq:tcoag}
\end{equation}
{where $n_{\rm H}$ is the number density of hydrogen atoms, $a$ is the maximum grain size, and $v_{rel}$ is the typical relative velocity of two dust grains.  We estimated $n_{\rm H}$ using the median best-fit $N(\rm H_{2})$ values in SMM-6: $2.5\times 10^{-22}$\,cm$^{-2}$ for \textit{Herschel}-only fitting, and $2.8\times 10^{22}$\,cm$^{-2}$ for the \textit{Herschel}+SCUBA-2 fitting.  We estimated a depth of SMM-6 of $9.0\times 10^{16}$\,cm, equal to $\sqrt{A}$, where $A$ is the area over which the properties described above are measured.  This led us to estimated $n_{\rm H_2}$ values of $2.8\times 10^{5}$\,cm$^{-3}$ from \textit{Herschel}-only fitting and $3.1\times 10^{5}$\,cm$^{-3}$ from \textit{Herschel}+SCUBA-2 fitting.  We therefore adopt $n_{\rm H} = 2\,n_{\rm H_{2}}\approx 6\times 10^{5}$\,cm$^{-3}$ for SMM-6.}

{Using $n_{\rm H} \approx 6\times 10^{5}$\,cm$^{-3}$ and $v_{rel}\sim 1$\,m\,s$^{-1}$, equation~\ref{eq:tcoag} suggests that the time required to grow grains to micron sizes in SMM-6 is $\sim 10^{7}$\,yr.  Since the CrA cloud contains many Class III protostars \citep[e.g.][]{peterson2011}, it must have an age of at least a few $\times 10^{6}$\,yr (cf. \citealt{evans2009}), but it is not clear that SMM-6 has existed in an undisturbed state for long enough for significant grain growth to have occurred.}

{This is further demonstrated by considering the ratio of $t_{coag}$ and the freefall time ($t_{ff}$) in SMM-6, again following \citet{chakrabarti2005}:}
\begin{equation}
    \frac{t_{coag}}{t_{ff}} \sim 2000\left(\frac{10^{5}\,{\rm cm}^{-3}}{n_{\rm H}} \right)^\frac{1}{2}\left(\frac{a}{5\,\mu{\rm m}}\right)\left(\frac{1\,{\rm m\,s}^{-1}}{v_{rel}}\right).
\end{equation}
{Using the same values for all terms as in the previous calculation, this relation suggests that the timescale for growth to micron-sized grains in SMM-6 is $>100\times t_{ff}$.  This would again require SMM-6 to be extremely long-lived for significant grain growth to have had time to occur within it.}
{While it is therefore difficult to physically motivate grain growth in SMM-6, the $\beta$ value that we measure in this region is nonetheless lower than the average across CrA from Planck data (see Section~\ref{sec:planck}, below).  This suggests that grain properties in and around the Coronet may differ from elsewhere in CrA. The dust properties of this nearby, well-resolved region may therefore be worthy of further investigation.}

\subsection{Comparison with Planck imaging}
\label{sec:planck}

We compared our results to the $\beta$ values determined across CrA by the \emph{Planck} Observatory \citep{planck2014}.  There is a significant discrepancy between the resolution of the \emph{Planck} $\beta$ maps (presented on 1\,arcmin pixels; \emph{Planck} has resolution $\sim 5$\,arcmin at 353\,GHz) and our own data.  Moreover, space-borne instrumentation is not subject to the atmospheric spatial filtering which restricts the size scales detectable by SCUBA-2.  This suggests that the \emph{Planck} data will on average trace lower-density material than SCUBA-2.  Despite this, we expect the values of $\beta$ inferred from \emph{Planck} data to provide a useful check on the results which we derive in this paper.

The \emph{Planck} $\beta$ map of the full CrA region is shown in Figure~\ref{fig:planck}.  $\beta$ varies significantly across CrA, exceeding 2 in CrA-C and CrA-E.  We find that the mean \emph{Planck} $\beta$ value over the full region observed with SCUBA-2 is $\beta_{Planck}^{all}=1.6 \pm 0.3$, while the mean value over the area where \emph{Herschel}-derived $N({\rm H}_{2}) \geq 10^{21}$\,cm$^{-2}$ (including 38 of our 39 cores) is $\beta_{Planck}^{\log N>21}=1.7 \pm 0.4$.  While the \emph{Planck} results suggest that the typical value of $\beta$ in CrA may be smaller than our chosen value of $\beta=2.0$, we nonetheless retain this value for purposes of comparison with previous studies, noting that the average \emph{Planck} value along high-column-density sightlines is consistent with $\beta=2.0$, and that the applicability of \emph{Planck}-derived values to cores on size scales of a few tens of arcseconds is unclear.

The mean value over the Coronet region outlined in Figures~\ref{fig:tempcompare} and \ref{fig:denscompare} is $\beta_{Planck}^{Coronet}=1.50\pm0.03$.  This is {consistent with} the value of {$\beta\approx 1.54$} which we find in SMM-6, suggesting that our SED fitting is producing an accurate value of $\beta$ in that region.

\section{Summary} \label{sec:summary}

In this paper, we have extracted a set of 39 starless and protostellar cores from the SCUBA-2 JCMT Gould Belt Legacy Survey data of the Corona Australis molecular cloud. We derived the properties for our catalogue of sources using the SCUBA-2 data, as well as a \textit{Herschel}-derived temperature map \citep{bresnahan2018}.

We showed that the empirical minimum volume density sensitivity of SCUBA-2 GBS data measured by \citet{dwt2016} holds across our sample of starless cores in CrA. We observed no correlation between the density and temperature of our cores, and find a population of higher-density prestellar cores with temperatures significantly higher than expected for such cores based on studies of other clouds. This matches previous \textit{Herschel} observations of prestellar cores in CrA \citep{bresnahan2018}.  The lack of correlation between core temperature and density is suggestive of significant external heating of cores in CrA.

We performed pixel-by-pixel SED fitting to SCUBA-2 and spatially-filtered \emph{Herschel} observations of the Coronet region.  We found {a low value of $\beta\approx1.54$} in the starless core SMM 6, {although we cannot confidently ascribe this to grain growth, since the timescale for grain coagulation is significantly longer than the freefall time in the region}.  Elsewhere in the Coronet cluster, some combination of $T-\beta$ fitting degeneracy, line-of-sight temperature variation and contamination of the SCUBA-2 850\um\ filter by the $^{12}$CO $J=3\to 2$ line prevent accurate determination of $\beta$, with values going as low as $\beta=0$ in our fitting.  We find that the Class I protostar IRS 2 has an apparently significant long-wavelength flux excess.  While we find {some evidence for variable $\beta$ values} in the Coronet region, comparison to \emph{Planck} data suggests that the canonical value $\beta=2.0$ is likely to be representative over the rest of the dense gas in CrA.  {This suggests that the Coronet region may be an excellent site for further investigation of how dust properties vary at high densities in star-forming regions.}

\section*{Data availability}

The raw SCUBA-2 data used in this paper are available from the JCMT archive at the Canadian Astronomy Data Centre (CADC; \url{https://www.cadc-ccda.hia-iha.nrc-cnrc.gc.ca/}) under project code MJLSG35, and the raw RxB data are available under project code M04AC05.  

The reduced data, and the complete set of images and the extended version of the catalogue shown in Appendix~\ref{sec:appendix_gs}, are available from [DOI to be inserted in proof].

\section*{Acknowledgements}
K.P. is a Royal Society University Research Fellow, supported by grant no. URF\textbackslash R1\textbackslash211322, and in an early stage of this project was supported by the Ministry of Science and Technology, Taiwan (Grant No. 106-2119-M-007-021-MY3).  D.J., H.K., J.D.F. and B.M. are supported by NRC Canada and by an NSERC Discovery Grant.

The James Clerk Maxwell Telescope (JCMT) is operated by the East Asian Observatory on behalf of The National Astronomical Observatory of Japan; Academia Sinica Institute of Astronomy and Astrophysics; the Korea Astronomy and Space Science Institute; the National Astronomical Research Institute of Thailand; Center for Astronomical Mega-Science (as well as the National Key R\&D Program of China with grant no. 2017YFA0402700). Additional funding support is provided by the Science and Technology Facilities Council (STFC) of the United Kingdom (UK) and participating universities and organizations in the UK, Canada and Ireland.

The JCMT has historically been operated by the Joint Astronomy Centre on behalf of the STFC of the UK, the National Research Council of Canada and the Netherlands Organisation for Scientific Research. Additional funds for the construction of SCUBA-2 were provided by the Canada Foundation for Innovation.  The data presented in this paper were taken under project code MJLSG35.

\textit{Herschel} is an ESA space observatory with science instruments provided by European-led Principal Investigator consortia and with important participation from NASA.

This research used: Starlink software \citep{currie2014}, supported by the East Asian Observatory; the services of the Canadian Advanced Network for Astronomy Research (CANFAR) which in turn is supported by CANARIE, Compute Canada, University of Victoria, the National Research Council of Canada, and the Canadian Space Agency; the facilities of the Canadian Astronomy Data Centre operated by the National Research Council of Canada with the support of the Canadian Space Agency; Astropy, a community-developed core Python package for Astronomy \citep{astropy2013, astropy2018, astropy2022}; the NASA Astrophysics Data System.

The authors wish to recognize and acknowledge the very significant cultural role and reverence that the summit of Mauna Kea has always had within the indigenous Hawaiian community. We are most fortunate to have the opportunity to conduct observations from this mountain.

\bibliographystyle{mnras}

\vspace{1cm}
$^{1}$Department of Physics and Astronomy, University College London, Gower Street, London WC1E 6BT, United Kingdom \\
$^{2}$Jeremiah Horrocks Institute, University of Central Lancashire, Preston, Lancashire, PR1 2HE, United Kingdom\\
$^{3}$NRC Herzberg Astronomy and Astrophysics, 5071 West Saanich Rd, Victoria, BC, V9E 2E7, Canada\\
$^{4}$Joint Astronomy Centre, 660 N. A`oh\={o}k\={u} Place, University Park, Hilo, Hawaii 96720, USA\\
$^{5}$Department of Physics and Astronomy, University of Victoria, Victoria, BC, V8P 5C2, Canada\\
$^{6}$Physics and Astronomy, University of Exeter, Stocker Road, Exeter EX4 4QL, United Kingdom\\
$^{7}$Vera C. Rubin Observatory, 950 N. Cherry Ave, Tucson, AZ, USA\\
$^{8}$Leiden Observatory, Leiden University, PO Box 9513, 2300 RA Leiden, The Netherlands\\
$^{9}$Max-Planck Institute for Astronomy, K{\"o}nigstuhl 17, 69117 Heidelberg, Germany\\
$^{10}$School of Physics and Astronomy, Cardiff University, The Parade, Cardiff, CF24 3AA, United Kingdom\\
$^{11}$Anton Pannekoek Institute for Astronomy, University of Amsterdam, PO Box 94249, 1090 GE Amsterdam, The Netherlands \\
$^{12}$Universit\'{e} de Montr\'{e}al, Centre de Recherche en Astrophysique du Qu\'{e}bec et d\'{e}partement de physique, C.P. 6128, succ. centre-ville, Montr\'{e}al, QC H3C 3J7, Canada\\
$^{13}$James Madison University, Harrisonburg, Virginia 22807, USA\\
$^{14}$The Department of Physics, Engineering \& Astronomy, Queen’s University, Stirling Hall, 64 Bader Lane, Kingston, ON K7L 3N6, Canada \\
$^{15}$SKA Observatory, Jodrell Bank, Lower Withington, Macclesfield SK11 9FT, UK\\
$^{16}$Department of Earth, Environment, and Physics, Worcester State University, Worcester, MA 01602, USA \\
$^{17}$Center for Astrophysics | Harvard \& Smithsonian, 60 Garden Street, Cambridge, MA 02138, USA \\
$^{18}$National Science Foundation, 2415 Eisenhower Avenue, Alexandria, VA 22314, USA \\
$^{19}$Medical Imaging and Biomedical Engineering, University College London, Gower Street, London WC1E 6BT, United Kingdom \\
$^{20}$Department of Physics and Astronomy, University of Waterloo, Waterloo, Ontario, Canada  N2L 3G1\\
$^{21}$Dept of Physics \& Astronomy, University of Manitoba, Winnipeg, Manitoba, R3T 2N2 Canada\\
$^{22}$Dunlap Institute for Astronomy \& Astrophysics, University of Toronto, 50 St. George St., Toronto ON M5S 3H4 Canada\\
$^{23}$Jodrell Bank Centre for Astrophysics, Alan Turing Building, School of Physics and Astronomy, University of Manchester, Oxford Road, Manchester, M13 9PL, United Kingdom\\
$^{24}$UK Astronomy Technology Centre, Royal Observatory, Blackford Hill, Edinburgh EH9 3HJ, UK\\
$^{25}$Institute for Astronomy, Royal Observatory, University of Edinburgh, Blackford Hill, Edinburgh EH9 3HJ, UK\\
$^{26}$Centre de recherche en astrophysique du Qu\'ebec et D\'epartement de physique, de g\'enie physique et d'optique, Universit\'e Laval, 1045 avenue de la m\'edecine, QC G1V 0A6, Canada\\
$^{27}$Astrophysics Group, Cavendish Laboratory, J J Thomson Avenue, Cambridge, CB3 0HE, United Kingdom\\
$^{28}$Kavli Institute for Cosmology, Institute of Astronomy, University of Cambridge, Madingley Road, Cambridge, CB3 0HA, United Kingdom\\
$^{29}$Department of Physics and Astronomy, McMaster University, Hamilton, ON, L8S 4M1, Canada\\
$^{30}$Department of Physics, University of Alberta, Edmonton, AB T6G 2E1, Canada\\
$^{31}$University of Western Sydney, Locked Bag 1797, Penrith NSW 2751, Australia\\
$^{32}$Dept. of Physical Sciences, The Open University, Milton Keynes MK7 6AA, United Kingdom\\
$^{33}$The Rutherford Appleton Laboratory, Chilton, Didcot, OX11 0NL, UK\\
$^{34}$National Astronomical Observatory of China, 20A Datun Road, Chaoyang District, Beijing 100012, China

\clearpage

\appendix

\section{Data}
\label{sec:appendix_data}

\renewcommand\thefigure{A\arabic{figure}}
\renewcommand\thetable{A\arabic{table}}

\begin{figure*}
\centering
%trim={<left> <lower> <right> <upper>}
\includegraphics[width=\textwidth]{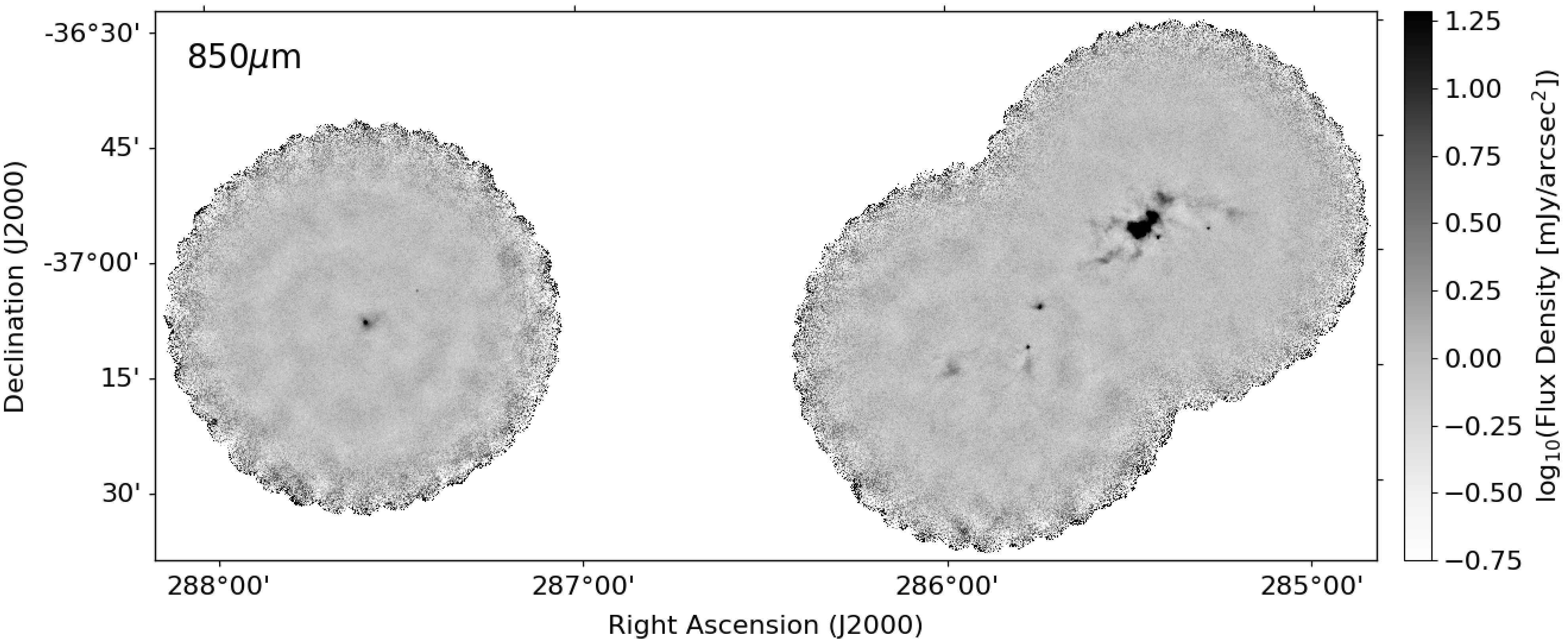}
\caption{The reduced SCUBA-2 850$\mu$m~flux density map, shown in logarithmic scaling.}
\label{fig:scubamap_850}
%\end{figure*}
%\begin{figure*}
%\centering
%trim={<left> <lower> <right> <upper>}
\includegraphics[width=\textwidth]{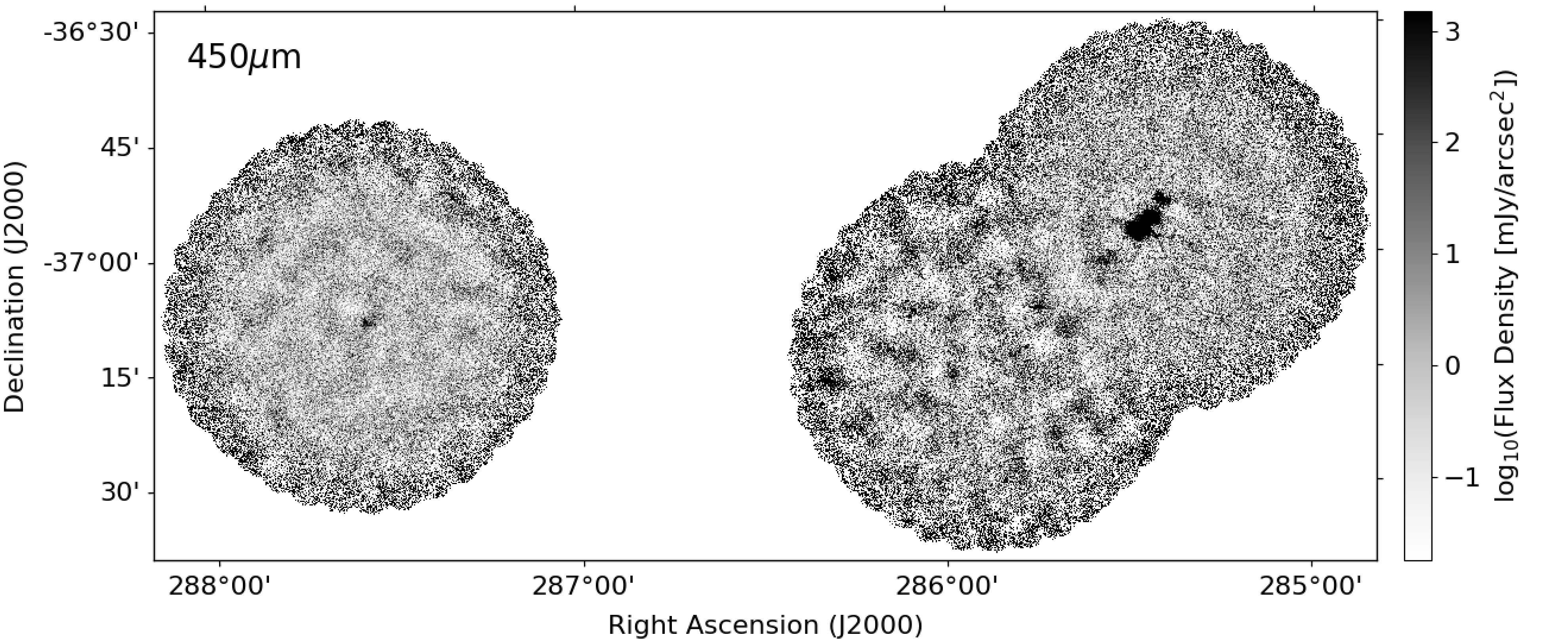}
\caption{The reduced SCUBA-2 450$\mu$m~flux density map, shown in logarithmic scaling.}
\label{fig:scubamap_450}
\end{figure*}

\FloatBarrier

\clearpage

\section{\textit{Getsources} output}
\label{sec:appendix_gs}

\renewcommand\thefigure{B\arabic{figure}}
\renewcommand\thetable{B\arabic{table}}

\begin{figure*}
	%\begin{center}
		%\begin{minipage}{1.0\linewidth}
		\centering
			%trim={<left> <lower> <right> <upper>}
		\resizebox{0.80\hsize}{!}{\includegraphics{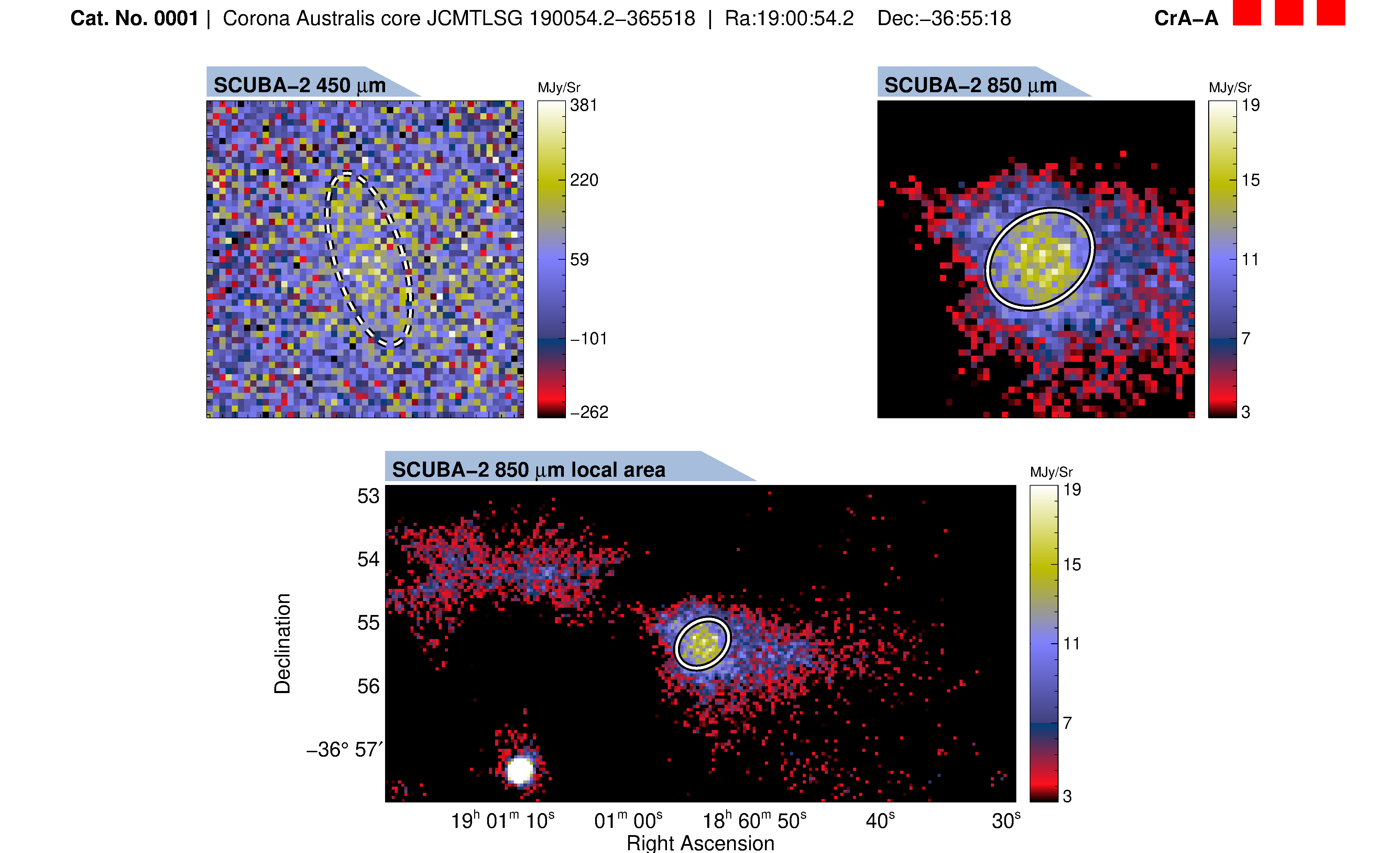}} 
		%\end{minipage}
	%\end{center}
\caption{Example \textit{getsources} output for a starless core. We show SCUBA-2 images at 450~$\mu$m~and 850~$\mu$m. We also show a local area cut-out using the 850$\mu$m~data. Ellipses represent the estimated major and minor FWHM sizes of the core at each wavelength. If a core is significantly detected at the respective wavelength, the line is solid, and is dashed otherwise. We provide a complete set of these images for our dense cores.}
     \label{jcmtcardexamplecore}%
\end{figure*}
\begin{figure*}
	%\begin{center}
		%\begin{minipage}{1.0\linewidth}
		\centering
			%trim={<left> <lower> <right> <upper>}
		\resizebox{0.80\hsize}{!}{\includegraphics{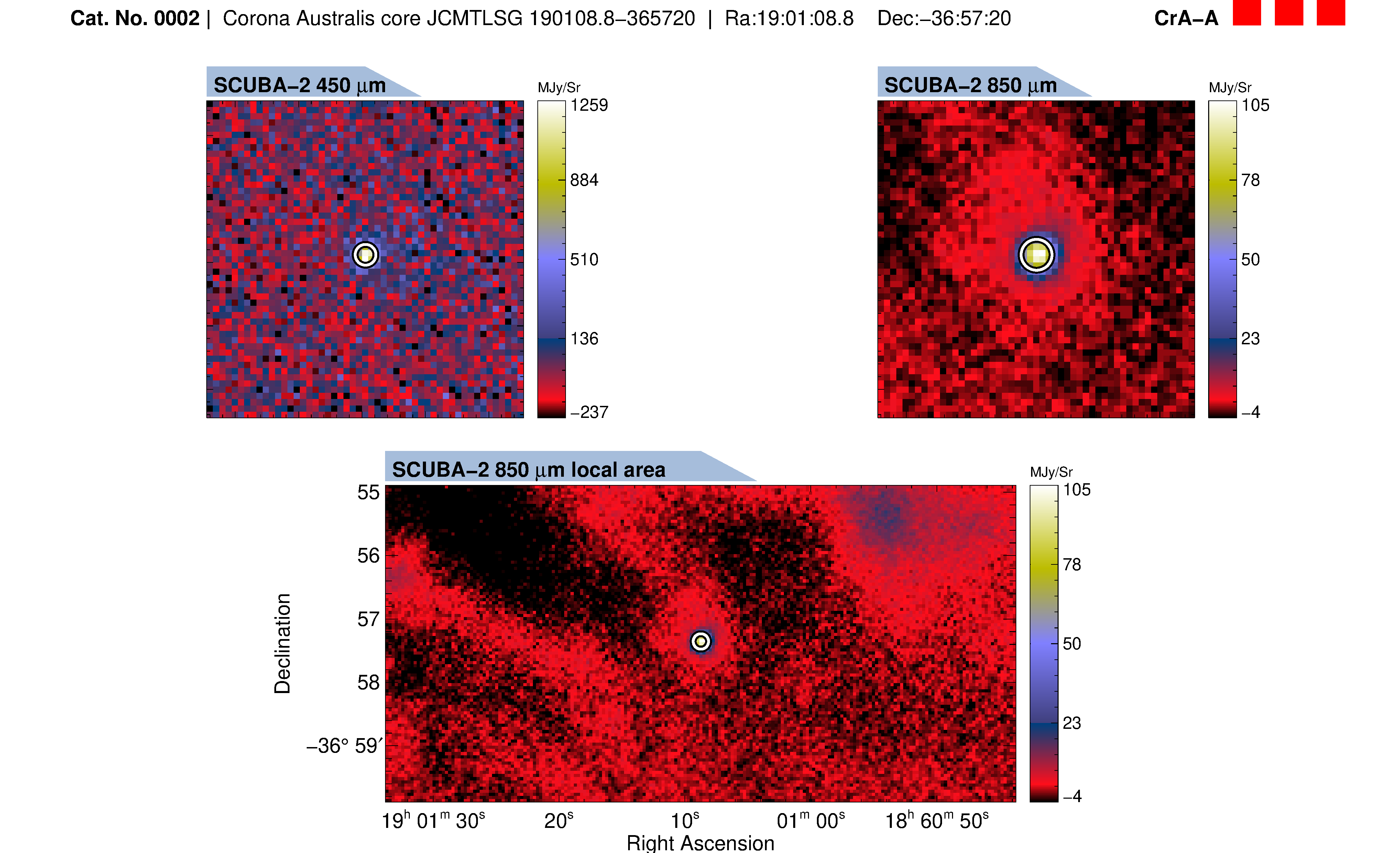}} 
		%\end{minipage}
	%\end{center}
\caption{As Figure \ref{jcmtcardexamplecore}, but for a protostellar core. The same image cut-outs are shown.}
     \label{jcmtcardexampleproto}%
\end{figure*}

\clearpage

%\begin{landscape}
%\begin{sidewaystable}[htbf]\scriptsize\setlength{\tabcolsep}{5.pt}
\begin{threeparttable}
\caption{Catalogue of dense cores identified in the JCMT GBLS maps of the Corona Australis molecular cloud (template, full catalogue only provided electronically).} 
\label{table:jcmtgbs_tab_obs_cat}
\renewcommand{\arraystretch}{1.2}
\begin{tabular}{ | c | c | r@{:}c@{:}l r@{:}c@{:}l | c r l c }
\hline\hline 
 C.No. & Source name & \multicolumn{3}{c}{R.A. (2000)} & \multicolumn{3}{c}{Dec. (2000)} & Sig$_{450}$ & \multicolumn{2}{c}{$S^\mathrm{peak}_{450}$} & $S^\mathrm{peak}_{450}/S_\mathrm{bg}$  \\ 
 & & \multicolumn{3}{c}{(h m s)} & \multicolumn{3}{c}{($^{\circ}$~\arcmin~\arcsec)} &  & \multicolumn{2}{c}{(Jy~beam$^{-1}$)} & \\ 
 (1) & (2) & \multicolumn{3}{c}{(3)} & \multicolumn{3}{c}{(4)} & (5) & \multicolumn{1}{c}{(6)} & \multicolumn{1}{c}{(7)} & (8) \\ 
\hline
  1 & 190054.1-365518 & 19 & 00 & 54.17 & -36 & 55 & 18.8 & 5 &    1.17e-01&   1.34e-01 &      0.47  \\
  2 & 190108.7-365720 & 19 & 01 & 08.78 & -36 & 57 & 20.4 & 18 &    2.92e+00&   2.91e-01 &     14.67  \\
  3 & 190121.6-365422 & 19 & 01 & 21.60 & -36 & 54 & 22.2 & -999 &    2.73e-01&   1.24e-01 &      1.17 \\
\hline\hline
\end{tabular}

\begin{tabular}{l r c c c c l r c}
\hline\hline 
   \multicolumn{2}{c}{$S^\mathrm{tot}_{450}$} & a$_{450}$ & b$_{450}$ & PA$_{450}$ &Sig$_{850}$ & \multicolumn{2}{c}{$S^\mathrm{peak}_{850}$} & $S^\mathrm{peak}_{850}/S_\mathrm{bg}$  \\ 
  \multicolumn{2}{c}{(Jy)} & (\arcsec) & (\arcsec) & ($^{\circ}$) & & \multicolumn{2}{c}{(Jy~beam$^{-1}$)} & \\ 
  \multicolumn{1}{c}{(9)} & \multicolumn{1}{c}{(10)} & (11) & (12) & (13) & (14) & \multicolumn{1}{c}{(15)} & \multicolumn{1}{c}{(16)} & (17) \\ 
\hline
 2.78e+00&   3.17e+00 &  84 &  32 & 163 & 16 &    8.08e-02&   1.35e-02 &     12.91  \\
 2.83e+00&   2.82e-01 &   9 &   9 &  41 & 103 &    5.62e-01&   1.17e-02 &     18.58 \\
 5.08e+00&   2.31e+00 &  49 &  45 &  84 & 9 &    2.24e-02&   1.15e-02 &      3.32  \\
\hline\hline
\end{tabular}

\begin{tabular}{ l r c c c c c c }
\hline\hline 
  \multicolumn{2}{c}{$S^\mathrm{tot}_{850}$} & a$_{850}$ & b$_{850}$ & PA$_{850}$ & \textsc{csar} & CuTEx & Core type \\
 \multicolumn{2}{c}{(Jy)} & (\arcsec) & (\arcsec) & ($^{\circ}$) & & & \\
\multicolumn{1}{c}{(18)} & \multicolumn{1}{c}{(19)} & (20) & (21) & (22) & (23) & (24) & (25) \\
\hline
  1.12e+00 &   1.88e-01 &  53 &  41 &  53 &  0 &    0 &  2 \\
  5.13e-01&   1.07e-02 &  14 &  14 &  72 &  2 &    2 &  3  \\
  8.16e-02&   4.19e-02 &  44 &  22 &  67 & 0 &    0 &  1  \\
\hline\hline
\end{tabular}

\setlength{\tabcolsep}{13.pt}
\begin{tabular}{ c c c c | }
\hline\hline 
  SIMBAD & WISE & \textit{Spitzer} & Comments \\
  & & & \\
 (26) & (27) & (28) & (29) \\
\hline
 [SHK2011b] 8 &  &  & no SCUBA-2 temperature/mass  \\
 V S CrA & J190108.61-365720.6 & S CrA & spatially unresolved at 850 micron                                                                   \\
    &  &  & no SCUBA-2 temperature/mass    \\
\hline\hline
\end{tabular}

\begin{tablenotes}
\item \scriptsize Catalog entries are as follows: 
{\bf(1)} Core number;
{\bf(2)} Core name $=$ JCMTGLS\_J prefix directly followed by a tag created from the J2000 sexagesimal coordinates; 
{\bf(3)} and {\bf(4)}: Right ascension and declination of core center; 
{\bf(5)} and {\bf(14)}: Detection significance from monochromatic single scales, in the 450$\mu$m and 850$\mu$m~maps, respectively. 
(NB: the detection significance has the special value of -999 when the core is not visible in clean single scales.); 
{\bf(6)}$\pm${\bf(7)} and {\bf(15)}$\pm${\bf(16)}: Peak flux density and its error in Jy/beam as estimated by \textsl{getsources};
{\bf(8)} and {\bf(17)}: Contrast over the local background, defined as the ratio of the background-subtracted peak intensity to the local background intensity ($S^{\rm peak}_{\rm \lambda}$/$S_{\rm bg}$); 
{{\bf(9)}$\pm${\bf(10)} and {\bf(18)}$\pm${\bf(19)}}: Integrated flux density and its error in Jy as estimated by \textsl{getsources}; 
{\bf(11)}--{\bf(12)} and {\bf(20)}--{\bf(21)}: Major \& minor FWHM diameters of the core (in arcsec), respectively, 
as estimated by \textsl{getsources}. (NB: the special value of $-1$ means that no size measurement was possible); 
{\bf(13)} and {\bf(22)}: Position angle of the core major axis, measured east of north, in degrees; 
{\bf(23)} `\textsl{CSAR}' flag: 2 if the \textsl{getsources} core has a counterpart detected by the \textsl{CSAR} source-finding algorithm \citep{kirk2013} within 6$\arcsec$ of its peak position,
equal to 1 if source found independently by \textsl{CSAR} within 6$\arcsec$ of \textsl{getsources} source, and 3 if the source is not already identified using the former criteria. However is located using the mask image produced by \textsl{CSAR}, 0 otherwise;
{\bf(24)} `\textsl{CuTEx}' flag: 2 if the \textsl{getsources} core has a counterpart detected by the \textsl{CuTEx} source-finding algorithm \citep{molinari2011} within 6$\arcsec$ of its peak position,
1 if no close \textsl{CuTEx} counterpart exists but the peak position of a \textsl{CuTEx} source lies within the FWHM contour of the \textsl{getsources} core in the 850$\mu$m~map, 0 otherwise;
{\bf(25)} Core type: 1=unbound starless, 2=prestellar, 3=dense core with embedded protostar; 
{\bf(26)} Closest counterpart found in SIMBAD, if any, up to 1$\arcmin$ from the {\it Herschel} peak position;
{\bf(27)} Closest WISE-identified YSO from the WISE all-sky YSO catalogue given by \citep{marton2015} within 6$\arcsec$ of the {\it Herschel} peak position, if any. When present, the WISE source name contained within the aforementioned catalogue is given;
{\bf(28)} Closest {\it Spitzer}-identified YSO from the \textit{Spitzer} Gould Belt Survey \citep{peterson2011} within 6$\arcsec$ of the SCUBA-2 850$\mu$m~peak position, if any. When present, the {\it Spitzer} source name contained within the aforementioned catalogue is given;
{\bf(29)} Comments may be \textit{no SCUBA-2 temperature/mass}, \textit{spatially unresolved at 850 micron}, \textit{reflection nebula?}.

\end{tablenotes}
\end{threeparttable}

\label{lastpage}

\end{document}